\documentclass[pre,aps,groupaddress,twocolumn,superscriptaddress]{revtex4}
\usepackage{epsfig,amsmath,graphicx,amssymb,overpic}
\usepackage{bm}
\usepackage{graphicx}
\usepackage{subfigure, xcolor}
\usepackage{soul} %
\usepackage{color, xcolor}
\soulregister{\cite}7

\def\be{\begin{equation}}
\def\ee{\end{equation}}
\def\bee{\begin{eqnarray}}
\def\ene{\end{eqnarray}}
\def\bes{\begin{subequations}}
\def\ees{\end{subequations}}

\newcommand{\PT}{{\cal PT}}

\begin{document}

\title{Dynamics of discrete solitons in the fractional discrete nonlinear Schr\"{o}dinger \\ equation with the quasi-Riesz derivative}

\author{Ming Zhong}
\affiliation{KLMM, Academy of Mathematics and Systems Science, Chinese Academy of Sciences,\\ Beijing 100190, China}
\affiliation{School of Mathematical Sciences, University of Chinese Academy of Sciences, Beijing 100049, China}

\author{Boris A. Malomed}
\affiliation{Department of Physical Electronics, School of Electrical Engineering, Faculty of Engineering,\\ Tel Aviv University,
Tel Aviv 69978, Israel}
\affiliation{Instituto de Alta Investigaci\'{o}n, Universidad de Tarapac\'{a},
Casilla 7D, Arica, Chile}

\author{Zhenya Yan}
\email{{\it Email:} zyyan@mmrc.iss.ac.cn (corresponding author)}
\affiliation{KLMM, Academy of Mathematics and Systems Science, Chinese Academy of Sciences,\\ Beijing 100190, China}
\affiliation{School of Mathematical Sciences, University of Chinese Academy of Sciences, Beijing 100049, China}

\baselineskip=12pt


\vspace{0.3in}

\begin{abstract}
\vspace{0.1in} We elaborate a fractional discrete nonlinear Schr\"{o}dinger
(FDNLS) equation based on an appropriately modified definition of the Riesz
fractional derivative, which is characterized by its L\'{e}vy index (LI).
This FDNLS equation represents a novel discrete system, in which the
nearest-neighbor coupling is combined with long-range interactions, that
decay as the inverse square of the separation between lattice sites. The
system may be realized as an array of parallel quasi-one-dimensional
Bose-Einstein condensates composed of atoms or small molecules carrying,
respectively, a permanent magnetic or electric dipole moment. The dispersion
relation (DR) for lattice waves and the corresponding propagation band in
the system's linear spectrum are found in an exact form for all values of
LI. The DR is consistent with the continuum limit, differing in the range of
wavenumbers. Formation of single-site and two-site discrete solitons is
explored, starting from the anti-continuum limit and continuing the analysis
in the numerical form up to the existence boundary of the discrete solitons.
Stability of the solitons is identified in terms of eigenvalues for small
perturbations, and verified in direct simulations. Mobility of the discrete
solitons is considered too, by means of an estimate of the system's
Peierls-Nabarro potential barrier, and with the help of direct simulations.
Collisions between persistently moving discrete solitons are also studied.
\newline


\end{abstract}

\maketitle

\vspace{0.3in}


\vspace{0.1in}

\section{Introduction}

The fractional derivative \cite{Uchaikin,Caputo} was originally developed as
a formal generalization of the classical integer derivative. The fractional
derivatives have found applications in a variety of fields of science and
engineering, such as non-Gaussian random processes, fluid mechanics, quantum
mechanics, optics, plasmas physics, Bose-Einstein condensates, biology,
environmental sciences, materials, control theory, electrical engineering,
signal and image processing, etc. (see, e.g., Refs.~\cite%
{fd1,fd2,fd3,St13,Lask1,Lask2,Lask3,fd-na,fd-np,Ku22}, books~\cite%
{nG,Lask4,b1,b2,b3,b4,b5} and reviews~\cite{PR1,PR2,fd-rev18}). There are
different formal definitions of the fractional derivatives. Many
mathematical works dealt with quite complex ones, such as the
Riemann-Liouville \cite{Uchaikin} and Caputo fractional\textit{\ }%
derivatives. The latter\ one is written as \cite{Caputo}%
\begin{equation}
D_{x}^{\alpha }\psi (x)=\frac{1}{\Gamma \left( 1-\{\alpha \}\right) }%
\int_{0}^{x}\frac{\psi ^{\left( n\right) }(\xi )}{\left( x-\xi \right)
^{\left\{ \alpha \right\} }}d\xi ,  \label{cap}
\end{equation}%
where $\alpha $ is the non-integer order of the derivative, $n\equiv \lbrack
\alpha ]+1$, $[\alpha ]$ is for the integer part of the fractional order, $%
\{\alpha \}\equiv \alpha -[\alpha ]$ is its fractional part, $\Gamma (\cdot
) $ is the Gamma-function, and $\psi ^{(n)}$ is the usual derivative of the
integer order $n$. Note that this and other definitions (see below)
represents the fractional derivative as an integral (nonlocal) operator,
rather than as a differential one.

In the context of physically relevant models, fractional derivatives had
originally appeared as the kinetic-energy operator in the framework of\
\textit{fractional quantum mechanics}, which was introduced by Laskin \cite%
{Lask1,Lask2,Lask3} (see also Ref.~\cite{GuoXu}) for nonrelativistic
particles which move, at the classical level, by \textit{L\'{e}vy flights }%
(random leaps), so that the mean distance $|x|$ of the particle, which moves
by one-dimensional (1D) ``flights" from the initial position, $x=0$, grows
with time $t$ as%
\begin{equation}
|x|\sim t^{1/\alpha },  \label{flight}
\end{equation}%
where the \textit{L\'{e}vy index} (LI) $\alpha $ takes values~\cite%
{Mandelbrot}: $0<\alpha \leq 2$. 
The limit value, $\alpha =2$, corresponds to the usual Brownian random walk,
while at $\alpha <2$ Eq.~(\ref{flight}) demonstrates that the L\'{e}vy
flights give rise to faster growth of $|x|$ at $t\rightarrow \infty $.

The derivation of the effective Schr\"{o}dinger equation as a result of the
quantization of a particle which moves, at the classical level, by L\'{e}vy
flight was performed in terms of the fundamental formalism which defines\
quantum mechanics by means of the Feynman's path integration. This formalism
represents the quantum dynamics of a particle as a result of the
superposition of virtual motions along all randomly chosen trajectories
(paths). The superposition is defined as the integral in the space of all
paths, $\sim \int \exp \left[ iS(\mathrm{path})\right] d(\mathrm{path})$,
where $S$ is the classical action corresponding to a particular path \cite%
{Feynman}. Laskin had developed this approach for the superposition of the
paths which correspond not to the usual Brownian random walks, but to chains
of L\'{e}vy flights. In the 1D setting, the result is the fractional linear
Schr\"{o}dinger (FLS) equation for wave function $\psi (x,t)$, which is written,
in the scaled form, as~\cite{Lask3,Longhi,Zhang15,Zhong16,ZZ16}
\begin{equation}
i\frac{\partial \psi }{\partial t}=\frac{1}{2}\left( -\frac{\partial ^{2}}{%
\partial x^{2}}\right) ^{\alpha /2}\psi +V(x)\psi ,\quad \alpha \in (0,2],
\label{FSE}
\end{equation}%
where $V(x)$ is the same real-valued external potential~\cite{Lask3,Longhi},
which appears in the usual Schr\"{o}dinger equation. The external potential
may also be chosen as a complex $\mathcal{PT}$-symmetric one~\cite{ZZ16}.

A result of the derivation of Eq.~(\ref{FSE}) by means of the path-integral
formalism is that the usual kinetic-energy operator, $-(1/2)\partial
^{2}/\partial x^{2}$, which corresponds to $\alpha =2$, is replaced in Eq.~(%
\ref{FSE}), at $\alpha <2$, by the fractional \textit{Riesz derivative} \cite%
{Riesz}. Its definition is more straightforward than that of the abstract
Caputo derivative (\ref{cap}), being composed as the juxtaposition of the
direct and inverse Fourier transforms ($x\rightarrow p\rightarrow x$) of the
wave function,
\begin{equation}
\left( \!\!-\frac{\partial ^{2}}{\partial x^{2}}\!\!\right) ^{\alpha
/2}\!\!\psi (x)\!=\!\frac{1}{2\pi }\!\!\int_{-\infty }^{+\infty
}dp|p|^{\alpha }\!\!\!\int_{-\infty }^{+\infty }d\xi e^{ip(x-\xi )}\psi (\xi
).\qquad  \label{Riesz derivative}
\end{equation}%
This definition implies that the action of the fractional derivative, $%
\left( -\partial ^{2}/\partial x^{2}\right) ^{\alpha /2}$, in the Fourier
space amounts to the straightforward multiplication: $\hat{\psi}%
(p)\rightarrow |p|^{\alpha }\hat{\psi}(p)$, where
\begin{equation}
\hat{\psi}(p)=\int_{-\infty }^{+\infty }\exp (-ipx)\psi (x)dx
\label{Fourier}
\end{equation}%
is the Fourier transform of the wave function. In definition (\ref{Riesz
derivative}), $\alpha $ is the same LI which determines the L\'{e}vy flights
as per Eq. (\ref{flight}).

The next step in the development of models originating from fractional
mechanics is to consider Bose-Einstein condensates (BECs) in an ultracold
gas of particles moving by L\'{e}vy flights \cite{review,review2,bec-fd}. It
is commonly known that, in the framework of the mean-field theory, the usual
BEC\ is governed by the Gross-Pitaevskii (GP) equation for the
single-particle wave function $\psi $, with the cubic term which represents
effects of inter-particle collisions \cite{Pit-Str}. While a systematic
derivation of the respective\textit{\ fractional} GP (FGP) equation remains
to be elaborated, its is natural to expect that it will take the form of the
GP equation with the kinetic-energy operator replaced by its fractional
counterpart defined as per Eqs. (\ref{Riesz derivative}). In the scaled
form, the equation will take the form of the respective fractional nonlinear
Schr\"{o}dinger (FNLS, alias FGP) equation~\cite%
{Huang08,Huang16,Zhang16,Yao18,Guo18,Li20,Xie19,Zeng20,Zhong23,Zhong24},%
\begin{equation}
i\frac{\partial \psi }{\partial t}=\frac{1}{2}\left( -\frac{\partial ^{2}}{%
\partial x^{2}}\right) ^{\alpha /2}\psi +V(x)\psi +\sigma |\psi |^{2}\psi ,
\label{FGPE}
\end{equation}%
where $\sigma =+1$ or $-1$ corresponds, respectively, to the repulsive or
attractive inter-particle collisions in the ultracold gas \cite%
{review,review2}, and $V(x)$ is a real-valued external potential, or complex
$\mathcal{PT}$-symmetric potential~\cite{Huang08,Xie19,Li20,Zhong23,Zhong24}%
. Available techniques make it possible to produce reliable numerical
solutions of the FNLS equations \cite{proceedings}.

Thus far, no experimental demonstration of the fractional quantum mechanics
or the corresponding fractional BEC has been reported. A more promising
physical realization of fractional Schr\"{o}dinger equation is suggested by
the commonly known similarity of the Schr\"{o}dinger equation and the
propagation equation for the amplitude of the optical field in the usual
case of paraxial diffraction, with time $t$ replaced, as the evolutional
variable, by the propagation distance, $z$ \cite{KA}. A scheme for the
realization of this possibility was proposed by Longhi~\cite{Longhi}, who
considered the transverse light dynamics in an optical cavity\ incorporating
two lenses and a phase mask, which is placed in the middle (Fourier) plane.
The lenses perform the direct and inverse Fourier transforms of the light
beam with respect to the transverse coordinate, which are ingredients of the
definition of the Riesz derivative in Eq.~(\ref{Riesz derivative}). Further,
according to Eq.~(\ref{Riesz derivative}), the action of the fractional
diffraction on the Fourier transform (\ref{Fourier}) of the optical
amplitude amounts to the phase shift,
\begin{equation}
\hat{\psi}\left( p\right) \rightarrow \hat{\psi}\left( p\right) \exp \left(
i|p|^{\alpha }Z\right) ,  \label{Z}
\end{equation}%
where $Z$ is the propagation distance which accounts for the fractional
diffraction. The corresponding differential phase shift of the Fourier
components is imposed by the phase mask placed in the midplane of the
optical cavity. The required mask can be created as a computer-generated
hologram \cite{Shilong}.

The setup which is outlined above provides a single-step transformation of
the optical beam. The continuous FNLS equation is then introduced as an
approximation for many cycles of circulation of light in the cavity,
assuming that each cycle introduces a small phase shift (\ref{Z}), while the
cubic term in the equation represents the Kerr nonlinearity of the optical
material.

An interesting ramification of the recently published theoretical works on
the topic of fractional media is the study of their discrete counterparts 
\cite{PRE95,CMP17,mathematical,Molina,Molina2,Molina-electro}, which are of nonlocal type. In particular,
Ref.~\cite{CMP17} provided a theoretical proof for the existence of
onsite and offsite fractional-order discrete solitary waves, encompassing a discussion on the Peierls-Nabarro potential barrier (PNPB). However, in those works, discrete
fractional derivatives were defined in a complex form, as a discrete version
of the Riemann-Liouville fractional derivative for the continuous field~\cite%
{Uchaikin}.  We here aim to define a new discrete fractional derivative in an
essentially simpler form, following the pattern of the continuous Riesz
derivative (\ref{Riesz derivative}). Our fractional discrete NLS (FDNLS)
equation is introduced in Section 2, where we also demonstrate its physical
realization, in the form of a system of GP equations for an array of
parallel chains of dipolar BECs. Some results, such as the dispersion
relation (DR) for the lattice waves, along with the corresponding band in
the system's linear spectrum (for all values of the respective L\'{e}vy
index (LI)), and strongly localized single- and double-site discrete
solitons in the anticontinuum limit (AL), are obtained in an analytical
form. Systematic numerical results for the structure and stability of
families of the discrete solitons, which originate, as the single- and
double-site ones in the AL approximation and are extended numerically up to
their existence boundaries, are reported in Section 3. In particular, we
find that the family of in-phase double-site discrete solitons is completely
unstable, while the family of out-of-phase ones includes a stable segment.
Mobility of the discrete solitons and collisions between mobile ones are
considered in Section 4. In particular, the mobility is predicted through
the consideration of the respective Peierls-Nabarro potential barrier
(PNPB). The work is concluded by Section 5.







\section{The fractional discrete NLS model}

\subsection{The formulation of the model with quasi-Riesz fractional
derivative}

A complex lattice function $u_{n}$ of the discrete coordinate, $n=0,\pm
1,\pm 2,\pm 3,...$, is represented by its Fourier transform, $U(k)$, which
is defined in interval $k\in \lbrack -\pi ,+\pi ]$ 
as a periodic function of real wavenumber $k$ in the Fourier space. The
direct and inverse relations between $u_{n}$ and $U(k)$ are%
\begin{equation}
u_{n}=\frac{1}{2\pi }\int_{-\pi }^{+\pi }U(k)e^{ikn}dk,\quad
U(k)\!=\!\sum_{n=-\infty }^{+\infty }u_{n}e^{-ikn}.  \label{inverse}
\end{equation}

In terms of the Fourier transform of continuous functions, the direct
counterpart of the Riesz fractional derivative, with L\'{e}vy index $\alpha
\in (1,2]$, of the complex discrete function $u_{n}$ can be derived as
follows:
\begin{equation}
\begin{array}{rl}
\displaystyle\left(-\frac{\widehat{\partial }^{2}}{\widehat{%
\partial}n^{2}}\right) ^{\alpha /2}\!\!u_{n} & \equiv \displaystyle\frac{1%
}{2\pi }\int_{-\pi }^{+\pi }dk\left( e^{ikn}|k|^{\alpha
}\!\!\!\!\sum_{m=-\infty }^{+\infty }\!\!u_{m}e^{-ikm}\right) \\
& =\displaystyle\frac{1}{\pi }\sum_{m=-\infty }^{+\infty }u_{m}\int_{0}^{\pi
}\cos \left( (m-n)k\right) \cdot k^{\alpha }dk \\
& =\displaystyle\sum_{\ell =-\infty }^{+\infty }D_{\ell }^{(\alpha
)}u_{n+\ell },%
\end{array}
\label{Riesz}
\end{equation}%
where the intersite coupling coefficients are%
\begin{equation}
\begin{array}{rl}
D_{\ell }^{(\alpha )}=D_{-\ell }^{(\alpha )} & =\displaystyle\frac{1}{\pi }%
\int_{0}^{\pi }\cos \left( \ell k\right) \cdot k^{\alpha }dk\vspace{0.1in}
\\
& = \displaystyle-\frac{\alpha }{\pi \ell }\int_{0}^{\pi }\sin \left(
\ell k\right) k^{\alpha -1}dk%
\end{array}%
\label{C}
\end{equation}%
(the latter expression, obtained by means of the integration by parts, is
more convenient for the subsequent analysis; obviously, the expression
(\ref{C}) vanishes, as it should, for all $\ell \neq 0$ at $\alpha =0$).
 Note that the notation for the novel fractional derivative of the
lattice function $u_{n}$, $\left( -\frac{\widehat{\partial }^{2}}{\widehat{%
\partial }n^{2}}\right) ^{\alpha /2}\!\!u_{n}$, given by Eq.~(\ref{Riesz}),
is symbolic, differing from the standard definition in the continuum limit (%
\ref{Riesz derivative}).  Similarly, the Fourier transform was also
utilized to give another discrete version for fractional
derivatives~\cite{Ta06,Ta062,Ta16}.

As a result, the definition given by Eq. (\ref{Riesz}) introduces a new
family of fractional discrete systems for $\alpha \leq 2$. In the limit case
of $\alpha =2$ (non-fractional diffraction), the discrete diffraction
operator includes coupling between non-neighboring sites, taking the form of%
\begin{equation}
\left( -\frac{\widehat{\partial }^{2}}{\widehat{\partial }n^{2}}\right)
u_{n}=\frac{\pi ^{2}}{3}u_{n}+2\sum_{\ell \in \mathbb{Z}\backslash
\{0\}}(-1)^{\ell }\ell ^{-2}u_{n+\ell }.  \label{alpha=2}
\end{equation}%
 The difference of Eq. (\ref{alpha=2}) from the
finite-difference operator of the coupling between nearest neighbors,
adopted in the commonly known DNLS equation (see Eq.~(\ref{DNLS}) below)
\cite{Ke09,Le08,dnls}, is explained by the fact that the usual operator is obtained by
first taking $\alpha \rightarrow 2$ in the continuum FNLS equation, and then
applying the straightforward discretization to the resulting NLS equation.
Operator (\ref{alpha=2}) is derived differently: first, the discretization
is applied to the continuum FNLS equation, and then the limit of $\alpha
\rightarrow 2$ is taken in it. Thus, we conclude that the two limit
procedures \emph{do not commute}, which explains the difference of the
operator (\ref{alpha=2}) from the straightforward finite difference. In fact,
the different discretization of the second-order derivative that we consider here is
a spectral format of the discretization~\cite{TLN}, which is thus a global
format, in contrast to finite differences.

For arbitrary $\alpha $, the integral in Eq. (\ref{C}) can be formally
expressed in terms of the incomplete Gamma-function, but this is useless for
practical applications, making it necessary to compute coefficients $%
D_{\ell}^{(\alpha )}$ numerically. For $\alpha <1$, the asymptotic form of
the coupling coefficients at $|\ell|\rightarrow \infty $ is%
\begin{equation}
D_{\ell}^{(\alpha )}\approx -\frac{\alpha \Gamma (\alpha )}{\pi
|\ell|^{\alpha+1 }}\sin \left( \frac{\pi }{2}\alpha \right) .  \label{asympt}
\end{equation}%
For $\alpha >1$, a similar formula is%
\begin{equation}
D_{\ell}^{(\alpha )}\approx \frac{(-1)^{\ell}\alpha}{\pi ^{2-\alpha }\ell^{2}%
}.  \label{asympt2}
\end{equation}

Expression (\ref{Riesz}) takes a much simpler form in the special case of $%
\alpha =1$:%
\begin{equation}
\left( -\frac{\widehat{\partial }^{2}}{\widehat{%
\partial}n^{2}}\right)
^{1/2}u_{n}=\frac{\pi }{2}u_{n}-\frac{2}{\pi }\sum_{\ell=-\infty }^{+\infty }%
\frac{u_{n+2\ell+1}}{\left(2\ell+1\right) ^{2}}.  \label{alpha=1}
\end{equation}%
Note that Eq.~(\ref{alpha=1}) includes the coupling of a given site, $n$, to
ones $n\pm 1$, $n\pm 3,n\pm 5$, ..., but not $n\pm 2,n\pm 4,n\pm 6$, etc.
The ``self-coupling" term $\frac{\pi }{2}u_{n}$ in Eq.~(\ref{alpha=1}) is
tantamount to a shift of the propagation constant.

Thus, based on the defined discrete quasi-Riesz fractional derivative (\ref%
{Riesz}), we introduce the novel fractional discrete NLS (FDNLS) equation
\begin{equation}
i\frac{du_{n}}{dt}=C\left( -\frac{\widehat{\partial }^{2}}{\widehat{%
\partial}n^{2}}\right) ^{\alpha /2}u_{n}-\sigma |u_{n}|^{2}u_{n},  \label{FNLSE-0}
\end{equation}%
whose actual form is%
\begin{equation}
i\frac{du_{n}}{dt}=C\sum_{\ell =-\infty }^{+\infty }D_{\ell }^{(\alpha
)}u_{n+\ell }-\sigma |u_{n}|^{2}u_{n},  \label{FNLSE}
\end{equation}%
where $C>0$ is the coefficient of the fractional discrete diffraction or the
linear coupling strength between adjacent sites of the lattice, and $\sigma
=+1$ or $-1$ represent the self-focusing or defocusing nonlinearity,
respectively. Note that the cubic nonlinear term may be replaced by
different ones~\cite{dnls, Le08}, such as the cubic-quintic term, $%
g_{1}|u_{n}|^{2}u_{n}+g_{2}|u_{n}|^{4}u_{n}$, the power-law nonlinearity, $%
|u_{n}|^{2p }u_{n}$, and the saturable expression, $%
u_{n}/(1+|u_{n}|^{2})$. Similar to its continuum counterpart, Eq.~(\ref%
{FNLSE-0}) conserves the Hamiltonian (energy) and power (norm),
\begin{equation}
H=\sum_{n}\left[ Cu_{n}^{\ast }\left( -\frac{\widehat{\partial }^{2}}{\widehat{%
\partial}n^{2}}\right) ^{\alpha /2}u_{n}-\frac{\sigma }{2}%
|u_{n}|^{4}\right] ,  \label{Ha}
\end{equation}%
\begin{equation}
P=\sum_{n}|u_{n}|^{2},  \label{Po}
\end{equation}%
where the star denotes the complex conjugate.
The conservation of the Hamiltonian and power corresponds, respectively, to
the invariance of Eq.~(\ref{FNLSE}) with respect to the time translation and
phase shift. Note that we can also consider the FDNLS Eq.~(\ref{FNLSE-0}) with a real or complex (e.g.,
$\PT$-symmetric) potential  $V(n)$,
\begin{equation}
i\frac{du_{n}}{dt}=C\left( -\frac{\widehat{\partial }^{2}}{\widehat{%
\partial}n^{2}}\right) ^{\alpha /2}u_{n}+V(n)u_n-\sigma |u_{n}|^{2}u_{n}. \label{FNLSE-0v}
\end{equation}
As $\sigma=0$, Eq.~(\ref{FNLSE-0v}) becomes the fractional discrete version of the continuous FLS Eq.~(\ref{FSE})
\begin{equation}
i\frac{du_{n}}{dt}=\frac{1}{2}\left(-\frac{\widehat{\partial}^{2}}
{\widehat{\partial}n^{2}}\right)^{\alpha/2}u_{n} +V(n)u_{n}.
\end{equation}

In particular, at $\alpha =1$, the cases of $\sigma =+1$ and $-1$ for Eq.~(%
\ref{FNLSE-0}) are mutually equivalent, as the latter case may be cast into
the former one by means of the well-known \textit{staggering transform}: the
substitution of%
\begin{equation}
u_{n}(t)\equiv (-1)^{n}q_{n}^{\ast }(t)e^{-iC\pi t},  \label{stagger}
\end{equation}%
where $\ast $ stands for the complex conjugate, in Eq.~(\ref{alpha=1})
replaces it (after the application of the complex conjugation) by the the
same equation for variables $q_{n}(t)$, but with $\sigma \rightarrow -\sigma
$. Thus, it is sufficient to consider only $\sigma =+1$ in the case of $%
\alpha =1$. The staggering transform does not apply to Eq. (\ref{FNLSE}) in
the general case of $\alpha \neq 1$.

For $\alpha =2$, the FDNLS equation (\ref{FNLSE-0}) reduces to a novel
integer-order DNLS equation
\begin{equation}
i\frac{du_{n}}{dt}=C\left( -\frac{\widehat{\partial }^{2}}{\widehat{%
\partial}n^{2}}\right) u_{n}-\sigma |u_{n}|^{2}u_{n}.  \label{FNLSE-02}
\end{equation}%
According to the definition of discrete derivative given by Eq.~(\ref{alpha=2}), the explicit form of Eq. (\ref{FNLSE-02})
is
\begin{equation}
i\frac{du_{n}}{dt}=C\left( \frac{\pi ^{2}}{3}u_{n}+2\!\!\sum_{\ell \in
\mathbb{Z}\backslash \{0\}}\frac{(-1)^{\ell}}{\ell^{2}}u_{n+\ell }\right) -\sigma
|u_{n}|^{2}u_{n},  \label{FNLSE-02v}
\end{equation}%
which, as stressed above, differs from the usual DNLS equation~\cite%
{dnls,Le08,Ke09}:
\begin{equation}
i\frac{du_{n}}{dt}=C(2u_{n}-u_{n+1}-u_{n-1})-\sigma |u_{n}|^{2}u_{n}.
\label{DNLS}
\end{equation}%
%
%
%
%

\subsection{\protect\vspace{0.1in} Discrete standing-wave solutions and
their stability}

In what follows below, we consider stationary solutions of Eq.~(\ref{FNLSE})
and their stability. First, we look for standing-wave solutions to Eq.~(\ref%
{FNLSE}) with $\sigma =1$ in the form of
\begin{equation}
u_{n}(t)=v_{n}e^{-i\omega t},  \label{omega}
\end{equation}%
with real frequency $\omega $ and stationary amplitudes $v_{n}$ satisfying
the steady-state equation
\begin{equation}
\left( \left\vert v_{n}\right\vert ^{2}+\omega \right) v_{n}=C\sum_{\ell
=-\infty }^{+\infty }D_{\ell }^{(\alpha )}v_{n+\ell }.  \label{SNLSE}
\end{equation}%
Note that $\omega <0$ can be scaled out from Eq.~(\ref{SNLSE}) by means of
the scaling transformation
\begin{equation}
v_{n}=\sqrt{-\omega }\hat{v}_{n},\quad C=-\omega \hat{C},
\end{equation}%
therefore we fix $\omega =-1$, keeping coupling constant $C$ as the control
parameter. In our numerical calculations, Eq. (\ref{Riesz}) is truncated
from $\ell =-100$ to $+100$.

\vspace{0.1in} Once standing-wave solutions of Eq.~(\ref{SNLSE}) have been
found, their linear stability is addressed by means of the Bogoliubov-de
Gennes linearized equations for small perturbations. To this end,
expressions for the perturbed solutions with an infinitesimal amplitude $%
\delta $ of the perturbations are substituted in Eq.~(\ref{FNLSE-0}) as
\begin{equation}
u_{n}(t)=e^{-i\omega t}\left[ v_{n}+\delta \left( a_{n}\mathrm{e}^{\lambda
t}+b_{n}^{\ast }\mathrm{e}^{\lambda ^{\ast }t}\right) \right]
\end{equation}%
with $\lambda \in \mathbb{C},\,\left( a_{n},b_{n}\right) \in \mathbb{C}%
^{2}[n],\,n\in \mathbb{Z}$, which gives rise to the eigenvalue problem
\begin{equation}
\left(
\begin{matrix}
L_{11} & L_{12}\vspace{0.1in} \\
-L_{12}^{\ast } & -L_{11}^{\ast }%
\end{matrix}%
\right) \!\!\left(
\begin{matrix}
a_{n}\vspace{0.1in} \\
b_{n}%
\end{matrix}%
\right) \!=\!i\lambda \left(
\begin{matrix}
a_{n}\vspace{0.1in} \\
b_{n}%
\end{matrix}%
\right)  \label{spectral}
\end{equation}%
with
\begin{equation}
\begin{array}{rl}
L_{11} & =\displaystyle C\left( -\frac{\widehat{\partial }^{2}}{\widehat{%
\partial}n^{2}}\right) ^{\alpha /2}-2|v_{n}|^{2}-\omega ,\vspace{0.1in} \\
L_{12} & =-v_{n}^{2}.%
\end{array}
\label{op}
\end{equation}%
The discrete soliton $u_{n}(t)$ is unstable if there exists $\lambda $ with
Re$(\lambda )>0$. Predictions for the linear stability are then verified by
direct simulations of the perturbed evolution, by dint of the fourth-order
Runge-Kutta method~\cite{At91}, monitoring the conservation of Hamiltonian (%
\ref{Ha}) and power (\ref{Po}) to validate the accuracy of the numerical
scheme.

\subsection{Implementation of the model in an array of dipolar BECs}

The fractional discrete nonlinear model which is based on Eq.~(\ref{FNLSE})
with the intersite coupling defined, for $\alpha =2$, by Eq.~(\ref{FNLSE-02}%
), can be implemented as an array of parallel quasi-one-dimensional
(quasi-1D) traps for the Bose-Einstein condensate (BEC)\ of atoms carrying a
permanent magnetic dipolar moment \cite{dipolar-BEC}, or the condensate of
small polar molecules with a permanent electric dipolar moment (cf. Ref.~%
\cite{HS}). Indeed, the energy density for the interaction between two
parallel 1D chains of dipoles separated by distance $L$ can be easily found,%
\begin{equation}
\mathcal{E}=-\frac{m_{1}m_{2}}{2\pi L^{2}}\sin \theta _{1}\cdot \sin \theta
_{2},  \label{mm}
\end{equation}%
where $m_{1,2}$ are dipolar densities in the two chains, $\theta _{1,2}$ are
fixed angles between the dipoles in each chain and its direction, and the
magnetic permeability or electric permittivity of vacuum is set to be $1$.

The fractional discrete model similar to the ones considered here
corresponds to the setup with the array of parallel quasi-1D traps uniformly
filled by the dipolar BEC, the discrete coordinate $n$ in Eq.~(\ref{FNLSE})
being the number of the trap in the array. Then, in the case of $m_{1}=m_{2}$
and $\theta _{1}=-\theta _{2}$, Eq.~(\ref{mm}) exactly corresponds to the
set of the coupling coefficients (\ref{alpha=2}) and (\ref{asympt2}).
Relation $\theta _{1}=-\theta _{2}$ implies that the dipolar polarization is
imposed by an external magnetic or electric field with a periodically
flipping spatial structure. One can also consider the system with unequal
dipole moments in the two chains, i.e., $m_{1}\neq m_{2}$ and/or $\theta
_{1}\neq \theta _{2}$. It will be described by a model similar to Eq.~(\ref%
{FNLSE}), but with the intersite interaction subjected to spatially periodic
modulation. On the other hand, Eqs.~(\ref{asympt}) and (\ref{alpha=1})
represent theoretical models which cannot be directly implemented by means
of the physical setting outlined here.

\begin{figure*}[t]
\centering
\vspace{-0.15in} {\scalebox{0.75}[0.75]{\includegraphics{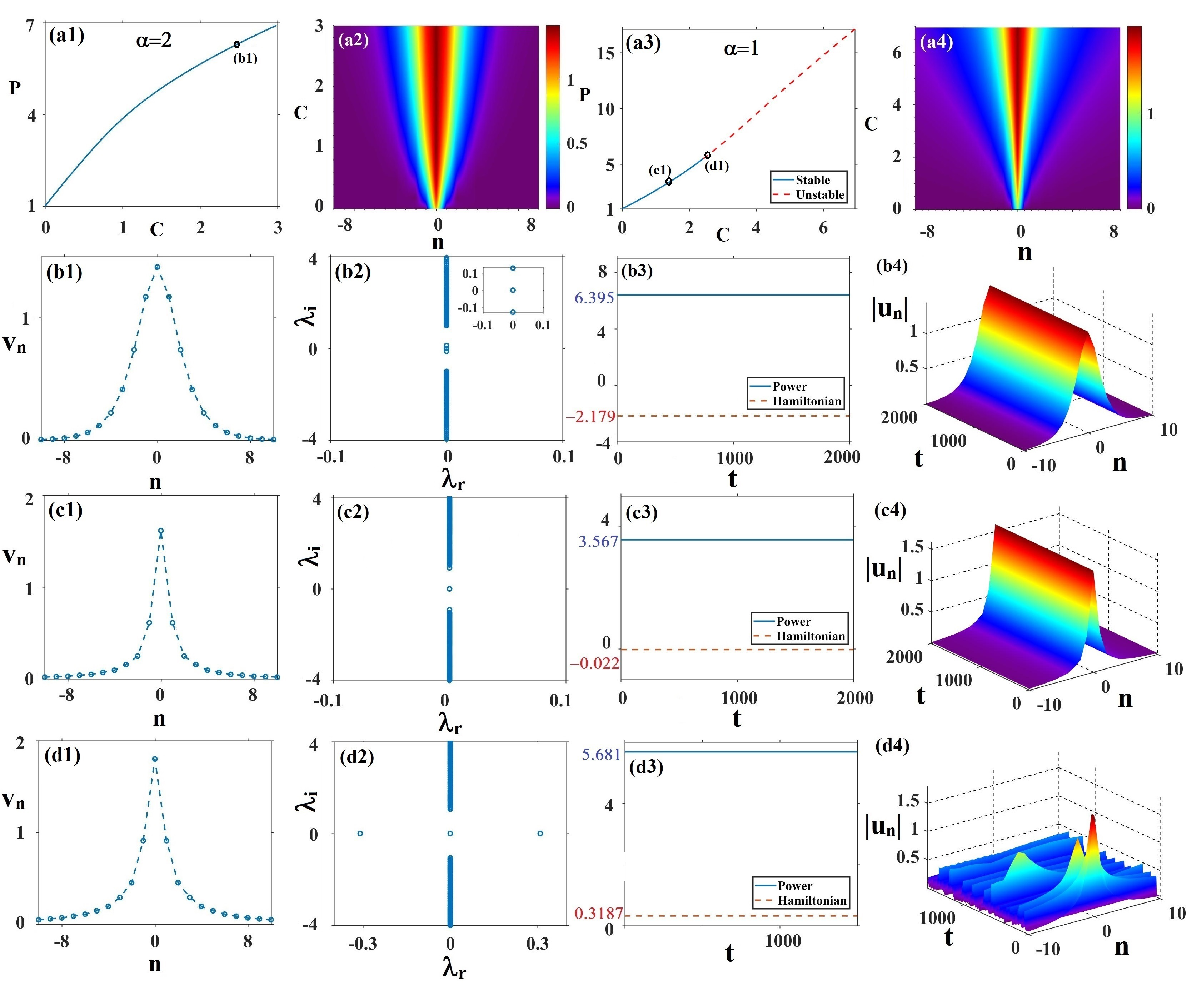}}}\hspace{%
-0.35in} \vspace{0.1in}
\caption{The single-site mode. The dependence of power $P$ on coupling
constant $C$ as produced by the numerical solution of Eq.~(\protect\ref%
{SNLSE}) with (a1) $\protect\alpha =2$ and (a3) $\protect\alpha =1$. The
 solid and dashed lines represent stable and unstable discrete solitons,
respectively. Labels (b1,c1,d1) correspond to discrete solitons displayed in
panels (b1,c1,d1). The variation of profiles $v_{n}$, following the increase
of coupling constant $C$ are displayed for $\protect\alpha =2$ (a2) and $%
\protect\alpha =1$ (a4). The second row, from left to right, displays the
profiles of discrete soliton (b1), linear spectrum (b2), power and
Hamiltonian in the course of the propagation (b3), and the nonlinear
dynamical behaviour (b4) for $\protect\alpha =2,C=2.5$.  The power and Hamiltonian are represented by solid and dashed lines, respectively. The third and fourth
rows are the same as the second one, but with parameters $\protect\alpha %
=1,C=1.5$ and $\protect\alpha =1,C=2.5$, respectively.}
\label{Single}
\end{figure*}

\begin{figure}[!t]
\centering
\vspace{-0.15in} {\scalebox{0.425}[0.425]{\includegraphics{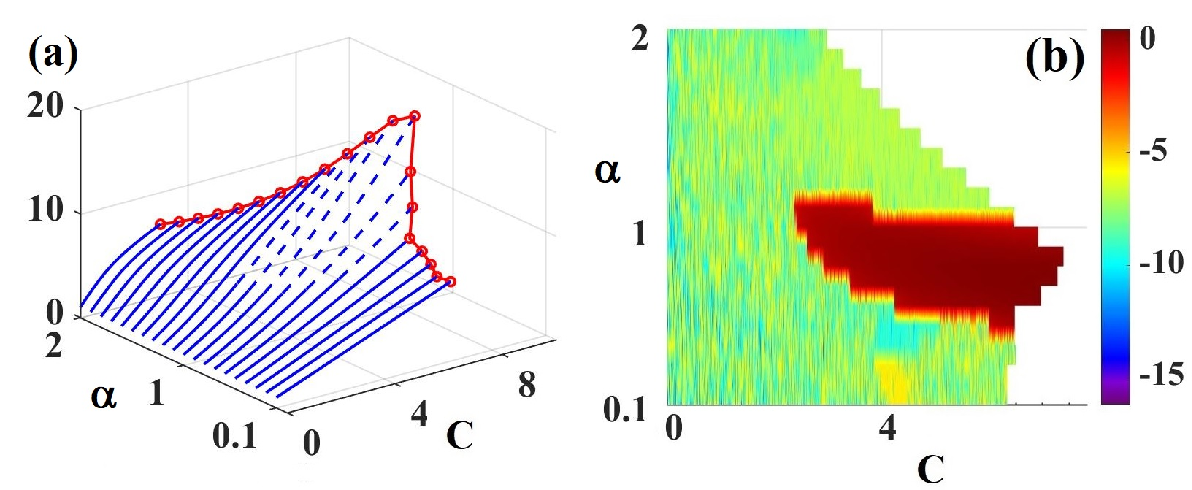}}}%
\hspace{-0.3in} \vspace{0.05in}
\caption{The single-site mode. (a) The dependence of power $P$ on coupling
constant $C$ and L\'{e}vy index $\protect\alpha $, where solid and dashed
lines denote stable and unstable segments, respectively. The red circle
denotes the boundary value of $C$ above which the discrete solitons do not
exist. (b) The largest real part $\protect\lambda _{r}$ of the linear
eigenvalues produced by the numerical solution of Eq.~(\protect\ref{spectral}%
) is charted in the $\left( C,\protect\alpha \right) $ plane. The boundary
of the chart corresponds to the chain of red circles in panel (a). The color
bar denotes values $\log (\max \{|\mathrm{Re}(\protect\lambda )|\})$, e.g.,
\textquotedblleft -5" corresponds to $\max \{|\mathrm{Re}(\protect\lambda %
)|\}=10^{-5}$ (actually, extremely small values imply stability of the
stationary mode).}
\label{Singlesu}
\end{figure}

\begin{figure*}[!t]
\centering
\vspace{-0.15in} {\scalebox{0.75}[0.75]{\includegraphics{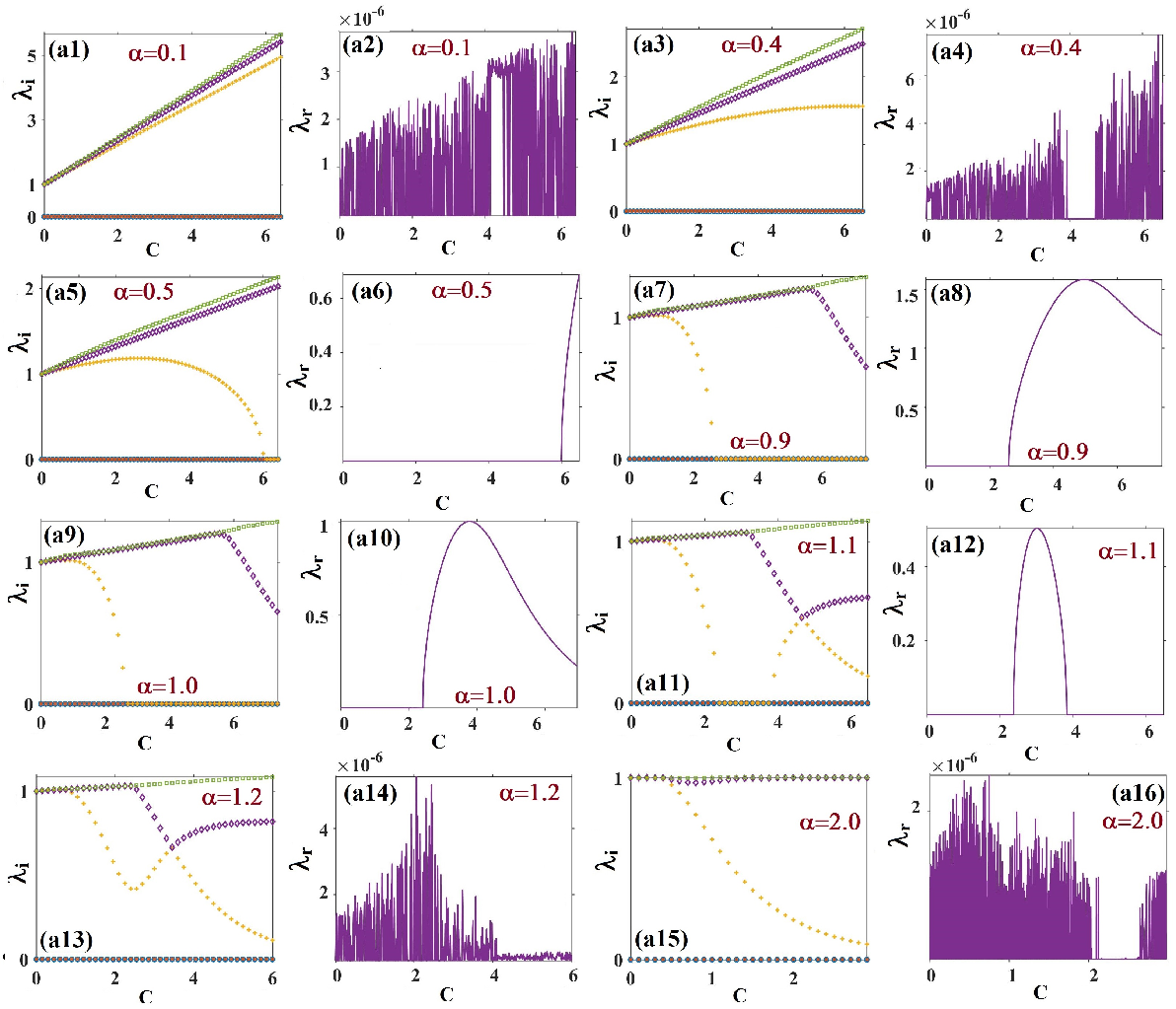}}}\hspace{%
-0.3in} \vspace{0.1in}
\caption{The single-site mode. Critical eigenvalues (including real and
imaginary parts) are produced by the numerical solution of Eq.~(\protect\ref%
{spectral}) with different L\'{e}vy indices $\protect\alpha $. The five eigenvalues with the smallest imaginary parts are labeled as follows: circles, crosses, pluses, diamonds, and squares. Several
spectra with positive imaginary parts and the corresponding real parts are
shown, \textit{viz}., (a1,a2): $\protect\alpha =0.1$; (a3,a4): $\protect%
\alpha =0.4$; (a5,a6): $\protect\alpha =0.5$; (a7,a8): $\protect\alpha =0.9$%
; (a9,a10): $\protect\alpha =1.0$; (a11,a12): $\protect\alpha =1.1$;
(a13,a14): $\protect\alpha =1.2$; (a15,a16): $\protect\alpha =2.0$.}
\label{SingleSP}
\end{figure*}

\subsection{The dispersion relation (DR) and spectral band for linear modes}

Fundamental characteristics of a discrete system is the dispersion relation
(DR) between real frequency $\omega $ and wavenumber $k$ of small-amplitude
lattice waves (\textquotedblleft phonons") governed by the linearized
system, and the spectral band (interval) of frequencies admitting the
propagation of the linear waves \cite{Ke09}. To identify the DR, the
solution of the linearized equation (\ref{SNLSE}) is looked for as $%
v_{n}\sim \exp \left( ikn\right) $, which yields%
\begin{equation}
\omega (k)=C\sum_{\ell =-\infty }^{+\infty }D_{\ell }^{(\alpha )}\cos \left(
\ell k\right) .  \label{om}
\end{equation}%
As it follows from definition (\ref{C}) of coefficients $D_{\ell }^{(\alpha
)}$ in terms of the Fourier integral, Eq. (\ref{om}) is tantamount to a
simple result,%
\begin{equation}
\omega (k)=C|k|^{\alpha },  \label{DR}
\end{equation}%
Notice that the discrete DR (\ref{DR}) is similar to the dispersion relation
in the continuous case, expect for the range of wavenumber $k$. For $\alpha
=2$, the DR of the integer-order discrete Eq.~(\ref{FNLSE-02v}) is $\omega
(k)=Ck^{2}$, which differs from the usual DNLS equation (\ref{DNLS}) whose
DR is $\omega (k)=2C(1-\cos (k))$.

The particular cases of Eq. (\ref{om}), \textit{viz}.,%
\begin{equation}
\omega (k=0)=C\sum_{\ell=-\infty }^{+\infty }D_{\ell}^{(\alpha )}=0,
\label{k=0}
\end{equation}%
and%
\begin{equation}
\omega (k=\pi )=C\sum_{\ell=-\infty }^{+\infty }(-1)^{\ell}D_{\ell}^{(\alpha
)}=C\pi ^{\alpha },  \label{k=pi}
\end{equation}
can be readily checked directly, using the known formulas
\begin{equation}
\sum_{\ell=1}^{\infty }\ell^{-1}\sin\left(\ell k\right) =(\pi -k)/2,
\label{sum1}
\end{equation}%
and
\begin{equation}
\sum_{\ell=1}^{\infty }(-1)^{\ell}\ell^{-1}\sin\left(\ell k\right) =-k/2,
\label{sum2}
\end{equation}%
which are valid for $0\leq k\leq \pi $ (note that the values produced by
Eqs. (\ref{sum1}) and (\ref{sum2}) at $k=0$ and $k=\pi $ are realized as
limit values of these expressions for $k\rightarrow 0+$ and $\pi
-k\rightarrow 0$, respectively). Thus, the lattice band, bounded by the
bottom and top points (\ref{k=0}) and (\ref{k=pi}) of the DR, is
\begin{equation}
0\leq \omega \leq C\pi ^{\alpha }.  \label{latt-band}
\end{equation}%
Discrete solitons may exist in the form of Eq. (\ref{omega}) at frequencies
which do not belong to the band. Indeed, it can be checked that intrinsic
frequencies of all the discrete solitons reported below are located outside
of the band.

Note also that DR (\ref{DR}) gives rise to a singularity of the group
velocity,
\begin{eqnarray}
V_{\mathrm{gr}}=d\omega /dk=C\alpha |k|^{\alpha -1}\mathrm{sgn}(k),
\end{eqnarray}
and of the respective dispersion coefficient, $d\omega /dk,$ at $%
k\rightarrow 0$ for $\alpha \leq 1$. This fact implies decay of propagating
lattice pulses in the latter case.

\subsection{The anti-continuum limit for discrete solitons}

It is well known that the~ anti-continuum (AC) limit, corresponding to $%
C\rightarrow 0$ in Eq. (\ref{FNLSE-0}), offers a convenient starting point
for constructing various solutions of DNLS equations \cite%
{Ma94,Au97,He99,Du00,Ka01,Al04,Pe05,Ke09}. In this limit, the sites $u_{n}$
are decoupled, hence Eq. (\ref{SNLSE}) amounts to a simple equation which
should be solved independently at each site:%
\begin{equation}
\left( \left\vert v_{n}\right\vert ^{2}-1\right) v_{n}=0.  \label{SNLSE2}
\end{equation}%
Obviously, its solutions are $v_{n}=0,$ or $v_{n}=\exp {\left( i\theta
_{n}\right) }$ with phase $\theta _{n}$ being a free parameter for each
site. Then, the crucial issue is what set of values of the phases at
different sites continues to a true solution at small nonzero values of the
coupling constant $C$. Indeed, according to the MacKay-Aubry's theorem~\cite%
{Ma94}, discrete solitons of Eq.~(\ref{FNLSE}) with $C=0$ and $v_n=\pm1,$ or
$0$ can be continued from the AC limit to finite values of $C$, provided
that the frequency $\omega $ does not resonate with linear modes of the
system.

While our fractional DNLS model (\ref{FNLSE-0}) is consistent with the usual
DNLS equation (\ref{DNLS}) in the AC limit, the linear spectra of our model
are strongly different, see below. In particular, substitution of $C=0$ in
Eq.~(\ref{spectral}) leads to the following eigenvalue problem:
\begin{eqnarray}
\left( 1-2v_{n}^{2}\right) a_{n}-v_{n}^{2}b_{n} &=&i\lambda a_{n},  \notag \\
v_{n}^{2}a_{n}-\left( 1-2v_{n}^{2}\right) b_{n} &=&i\lambda b_{n},
\label{SPAC}
\end{eqnarray}%
where $n\in \mathbb{Z}$, and the limit solution can be chosen to be
real-valued.\ Supposing that it includes $N$ excited sites with $v_{n}=\pm 1$
and $v_{n}=0$ at other sites in the AC limit, a direct solution of Eq.~(\ref%
{SPAC}) produces $N$ pairs of zero eigenvalues, together with an infinite
number of eigenvalues $\lambda =\pm i$. According to the discrete
Sturm-Liouville theorem~\cite{Le92} and regular perturbation theory~\cite%
{Ho12}, the infinite set of the eigenvalues is converted into the continuous
spectrum near $\lambda =\pm i$ at small finite values of $C$, without
producing any unstable eigenvalues with $\text{Re}\left( \lambda \right) >0$%
. As for the $N$ pairs of zero eigenvalues, the invariance with respect to
the phase shift keeps one pair of exact zero eigenvalues at $C>0$. The
critical issue for the stability analysis is to identify $N-1$ eigenvalue
pairs which had $\lambda =0$ at $C=0$, and become nonzero at $C>0$.

\section{Discrete solitons: formation and stability}

In what follows, the existence and stability of discrete solitons are
studied numerically. We mainly focus on the discrete solitons coded by a
sequence $\beta _{n}=\{0,+,-\},n\in \mathbb{Z}$ excited on $N=1$ or $2$
sites~(cf. Ref. \cite{Al04}), where $\beta _{n}=\{0,+,-\}$ indicates the
values $v_{n}=0$ or $\pm 1$ in the AC limit.

\subsection{Single-site excitations}

\begin{figure*}[!t]
\centering
\vspace{-0.15in} {\scalebox{0.8}[0.8]{\includegraphics{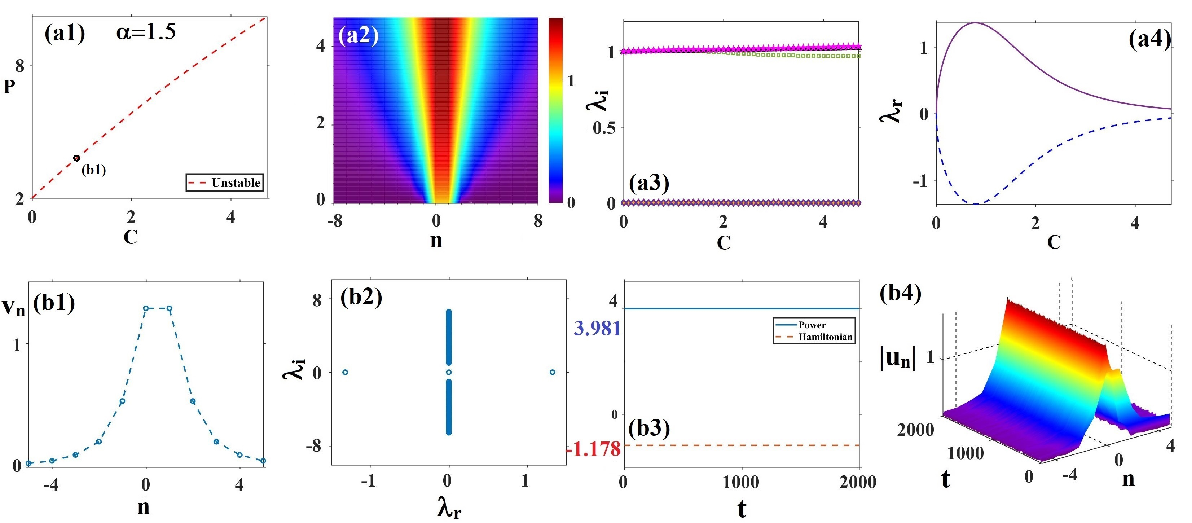}}}\hspace{-0.3in}
\vspace{0.1in}
\caption{Unstable two-site in-phase modes at $\protect\alpha =1.5$ and $C=1$%
. (a1) Power $P$ vs. the coupling constant $C$, as produced by the numerical
solutions of Eq.~(\protect\ref{SNLSE}). Label (b1) corresponds to the
discrete solitons displayed in panel (b1). (a2) The variation of profiles $%
v_{n}$, following the increase of $C$. (a3,a4) Critical eigenvalues
(including their real and imaginary parts), produced by the numerical
solution of Eq.~(\protect\ref{spectral}). The seven eigenvalues with the smallest imaginary parts are labeled in panel (a3) as follows: circles, crosses, plus signs, diamonds, squares,  upward-pointing triangles, and five-pointed stars.
The solid and dashed lines in panel (a4) represent unstable eigenvalues branching out from the original.
 The bottom row of panels: the
profile of the discrete soliton (b1), the spectrum of eigenvalues for small
perturbations about the discrete soliton (b2), the evolution of the discrete
soliton's power and Hamiltonian in the course of the perturbed propagation
(b3), and the corresponding evolution of the weakly unstable discrete
soliton (b4). }
\label{Two}
\end{figure*}

\begin{figure*}[!t]
\centering
\vspace{-0.15in} {\scalebox{0.7}[0.68]{\includegraphics{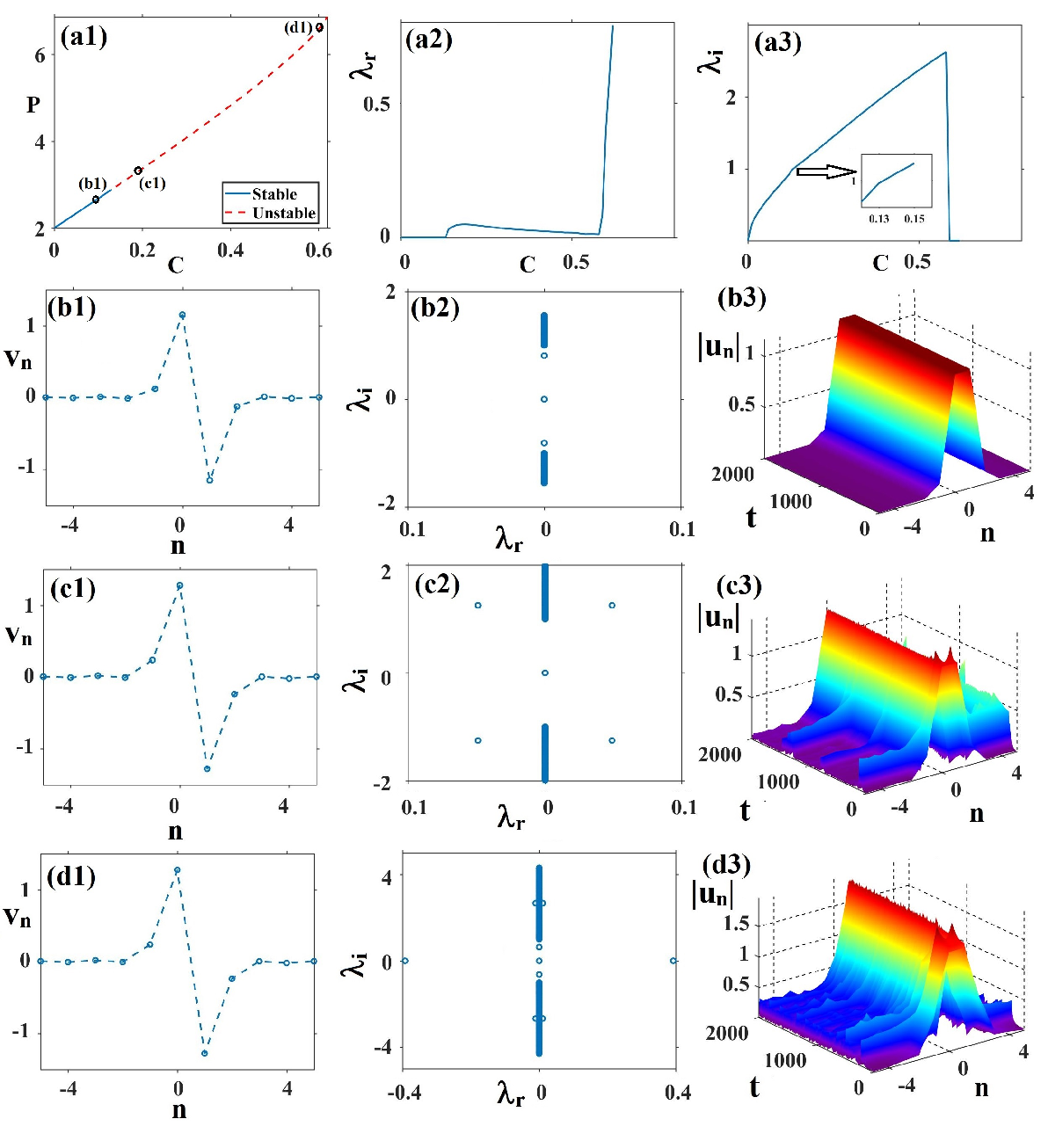}}}\hspace{%
-0.3in} \vspace{0.1in}
\caption{Two-site out-of-phase modes at $\protect\alpha =1.5$. (a1) Power $P$
vs. the coupling constant $C$, as produced by the numerical solutions of
Eq.~(\protect\ref{SNLSE}). The solid and dashed lines represent stable and
unstable discrete solitons, respectively. Labels (b1,c1,d1) correspond to
the discrete solitons displayed in panels (b1,c1,d1), respectively. (a2,a3)
Critical eigenvalues (including real and imaginary parts) produced by the
numerical solution of Eq.~(\protect\ref{spectral}). The second row of panels
display the profile of a discrete soliton for $C=0.1$: (b1), the
corresponding spectrum of perturbation eigenvalues (b2), and the evolution
of the discrete soliton (b3). The third and fourth rows show the same as the
second one, but for $C=0.2$ and $C=0.6$, respectively.}
\label{Dipole}
\end{figure*}

\begin{figure*}[!t]
\centering
\vspace{-0.15in} {\scalebox{0.68}[0.68]{\includegraphics{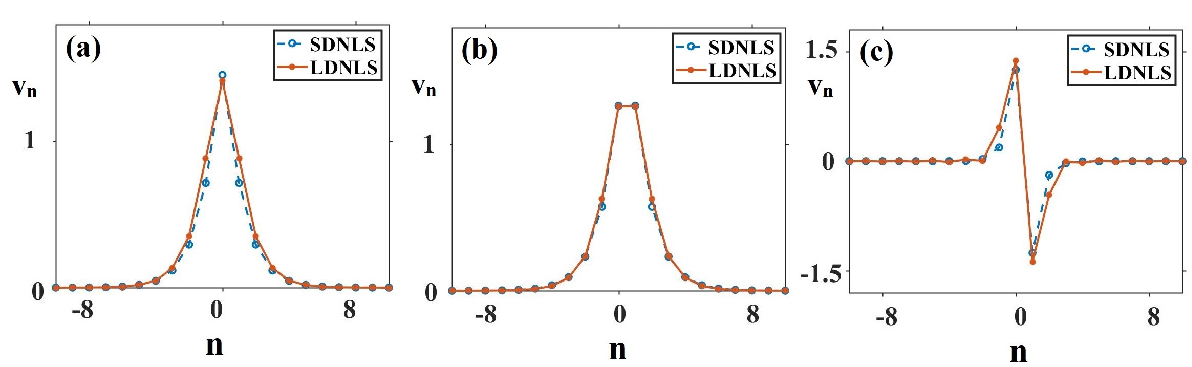}}}\hspace{-0.3in}
\vspace{0.1in}
\caption{ Comparison of different discrete soliton profiles for
DNLS equation (\protect\ref{FNLSE-02v}) with long-range interactions (labeled as
LDNLS  with solid line), and for the usual short-range DNLS equation  (\protect\ref{DNLS}) (labeled as
SDNLS  with dashed line): (a) and (b) represent, respectively, the onsite- and
offsite-centered solitons at $C=1$; (c) represents twisted solitons at $%
C=0.2 $. }
\label{Comp}
\end{figure*}

\begin{figure*}[!t]
\centering
\vspace{-0.15in} {\scalebox{0.6}[0.6]{\includegraphics{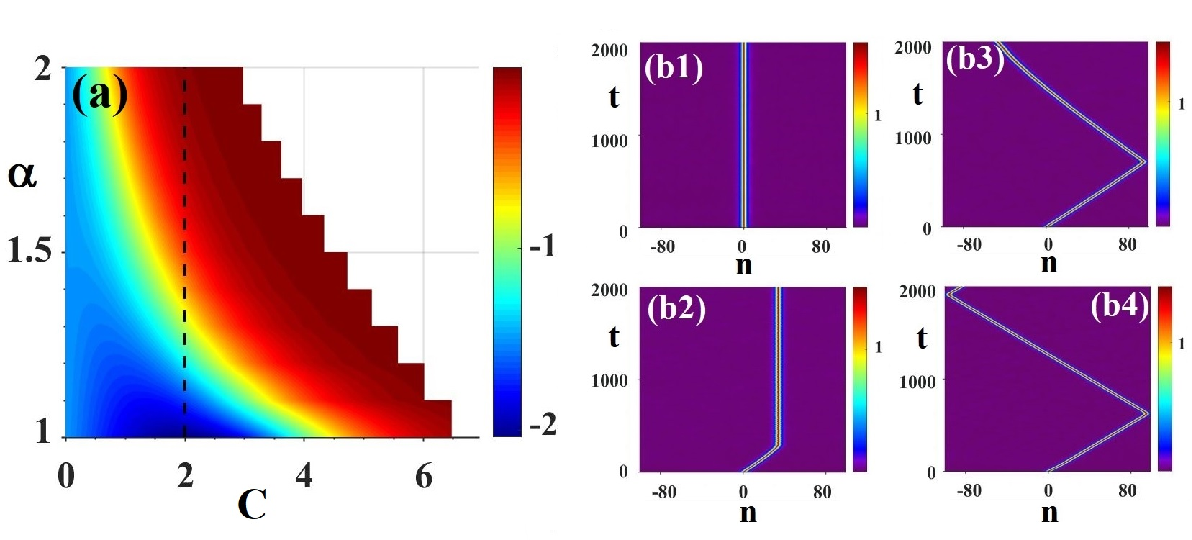}}}\hspace{%
-0.3in} \vspace{0.05in}
\caption{(a) The magnitude of the Peierls-Nabarro barrier, $\Delta G$,
estimated as per Eq.~(\protect\ref{FreeD}), is shown in the $\left( C,%
\protect\alpha \right) $ plane. The evolution of the discrete soliton
initiated by kick $K=\protect\pi /64$ (see Eq. (\protect\ref{kick})), for $%
C=0.2$ and $\protect\alpha =1.5$ (b1), $\protect\alpha =1.7$ (b2), $\protect%
\alpha =1.8$ (b3), and $\protect\alpha =2.0$ (b4).}
\label{Moving}
\end{figure*}

We start the analysis with discrete solitons originating from the
single-site ($N=1$, alias onsite-centered) excitation in the AC limit.
Generic examples of discrete solitons for L\'{e}vy indices $\alpha =2$ and $%
\alpha =1$ are presented in Fig.~\ref{Single}. Results for the fundamental
discrete solitons produced by the numerical solution of Eq.~(\ref{SNLSE})
with $\alpha =2$ and $\alpha =1$ are summarized in Figs.~\ref{Single}(a1)
and (a3), respectively, in the form of dependence $P(C)$. It is worthy to
note that, unlike the case of the nearest-neighbor interaction~\cite%
{Au97,Ke09}, the presence of the long-range coupling here does not allow the
single-site discrete soliton to proceed from the AC limit to the continuum
limit, even though the shape of the mode is very close to the one in the
continuum form. This situation is similar to that known for discrete
solitons in the recently considered fractional Frenkel-Kontorova model~\cite%
{Ca23}. The critical value of coupling constant $C$ above which no discrete
soliton can be found is $C_{\mathrm{cr}}(\alpha =2)=2.9820$ and $C_{\mathrm{%
cr}}(\alpha =1.0)=6.9390$. The variation of profiles $v_{n}$, following the
increase of coupling constant $C$ are displayed at Figs.~\ref{Single}(a2)
and (a4), respectively. One can observe that the discrete soliton with $%
\alpha =2$ is localized stronger than its counterparts with $\alpha =1.0$,
which is a manifestation of the existence of power-law tails in the case of
the long-range interaction~\cite{Ca23,Fl98}, similar to the continuum
FNLS~equation \cite{proceedings}.

We have found that all discrete solitons are stable for $\alpha =2$, which
is the same as in the case of the nearest-neighbor interaction~\cite%
{Au97,Ke09}. However, for $\alpha =1$ the discrete solitons become unstable
above the respective critical value of the coupling constant.
The instability is related to the fact that, in the framework of the
continuum FNLS\ equation with $\alpha =1$ the solitons are destabilized by
the critical collapse~\cite{proceedings}. Several profiles of discrete
solitons are displayed in Figs.~\ref{Single}(b1,c1,d1). It is seen in Fig.~%
\ref{Single}(b2) that the discrete soliton with $\alpha =2,C=2.5 $ displayed
in Fig.~\ref{Single}(b1) is a linearly stable mode. The spectrum of the
eigenvalues, displayed in the latter panel, contains a double zero
eigenvalue, which is due to the invariance with respect to the phase shift.
The simulated evolution, displayed in Fig.~\ref{Single}(b4) with random
noise added to the initial condition, verifies the stability predicted by
the linear spectrum, the accuracy of the simulations being confirmed by the
conservation of the Hamiltonian and power in Fig.~\ref{Single}(b3).

Next, we examine the stability of the discrete soliton for $\alpha
=1.0,~C=1.5$, as shown in Fig.~\ref{Single}(c1). The linear spectra and
direct simulations indicate that the discrete soliton is completely stable,
see Figs.~\ref{Single}(c2,c3,c4)). Lastly, we consider the discrete soliton
with $\alpha =1.0,C=2.5$ displayed in Fig.~\ref{Single}(d1). One observes in
Fig.~\ref{Single}(d2) that this discrete soliton is linearly unstable, with
the unstable eigenvalues bifurcating from the bottom of the continuous
spectrum, cf. Refs. \cite{Ka01,Pe05,Ke09}. The direct simulations confirm
the instability of this discrete soliton, see Figs.~\ref{Single}(d3,d4).

A more detailed study of the dependence of power $P$ on the coupling
constant $C$ and L\'{e}vy index $\alpha $ is presented in Fig.~\ref{Singlesu}%
(a). It is seen in Fig.~\ref{Singlesu}(a) that the critical value of the
coupling constant $C$, which is the existence boundary for the discrete
solitons, decreases with the growth of $\alpha $ at $\alpha \geq 0.7$, while
the results for $\alpha <0.7$ may be affected by the finite size of the
underlying lattice and boundary conditions. The largest real part of the
linear eigenvalue produced by Eq.~(\ref{spectral}) is shown in the $\left(
C,\alpha \right) $ plane in Fig.~\ref{Singlesu}(b), which shows that the
discrete solitons are unstable only when $0.5\leq \alpha \leq 1.1$, with $C$
exceeding a certain threshold value.

In an attempt to explore reasons for the generation of the instability, we
consider in detail the critical eigenvalues, which determine the stability,
by means of the numerical solution of Eq.~(\ref{spectral}), following the
variation of L\'{e}vy index $\alpha $. It is relevant to mention that, in
the continuum limit (in the model with the fractional and non-fractional
diffraction alike), in addition to the above-mentioned invariance with
respect to the phase shift, the invariance against the spatial translation
gives rise to an additional pair of zero eigenvalues, hence their total
number is four. We present some generic examples in Fig.~\ref{SingleSP}.
When $0.1\leq \alpha <0.4$, there is a single pair of zero eigenvalues, and
no eigenvalues bifurcate from the bottom edge of the continuous spectrum,
see Fig.~\ref{SingleSP}(a1). Further, no unstable eigenvalues appear with
the increase of the coupling constant $C$, see Fig.~\ref{SingleSP}(a2). When
$\alpha =0.4$, one observes in Fig.~\ref{SingleSP}(a3) that the translation
eigenvalues bifurcate from the continuous spectrum with the increase of $C$.
They move along the imaginary axis, maintaining the linear stability, see
Fig.~\ref{SingleSP}(a4). A phenomenon presented in Fig.~\ref{SingleSP}(a5),
which is similar to that produced by the DNLS equation~with saturable
nonlinearity \cite{St04,Ha04,Me06}, is that the translation eigenvalue
emerges from the band edge of the continuous spectrum at $\alpha =0.5$. It
vanishes at $C=6.01$, then becoming real and expanding along the real axis,
which results in instability, as seen in Fig.~\ref{SingleSP}(a6). Another
critical value of the L\'{e}vy index is $\alpha =0.9$, in which case another
pair of eigenvalues bifurcate from the lower edge of the continuous
spectrum, see Figs.~\ref{SingleSP}(a7,a8). Simultaneously, the translational
eigenvalues (real ones) decrease with the generation of the above-mentioned
pair. Similar results are observed at $0.9\leq \alpha <1.1$, see Figs.~\ref%
{SingleSP}(a9,a10). An intriguing phenomenon occurs at $\alpha =1.1$ is
displayed in Fig.~\ref{SingleSP}(a11), where the translation eigenvalue
first bifurcates from the continuous spectrum, then expands along the real
axis, and finally decreases due to the presence of other separate
eigenvalues. After returning to the origin, this eigenvalue spreads out
along the imaginary axis, and collides with the point spectrum branching off
from the continuous spectrum mentioned above. The collision is followed by
the motion of the two eigenvalues in the opposite directions. Note that such
a collision does not create unstable eigenvalues~\cite{Jo00}. The
corresponding changes in the real part of the eigenvalues are visualized in
Fig.~\ref{SingleSP}(a12). Note that the motion of the two eigenvalues in the
opposite direction after the collision is inevitable, as can be deduced from
the spectral distribution in the continuum limit. When $\alpha $ increases
to $1.2$, before the translation eigenvalues hit the origin, another pair of
eigenvalues branches off from the continuous spectrum, therefore no unstable
eigenvalues are generated. Similar collision phenomena are observed in Figs. %
\ref{SingleSP}(a13,a14). When $\alpha $ increases to $2$, as shown in Figs.~%
\ref{SingleSP}(a15,a16), no collisions are observed, and the eigenvalues
branching off from the continuous spectrum undergo a recurrence in which the
translation eigenvalues extend all the way along the imaginary axis without
touching the origin. This scenario is also different from those known in the
case of the nearest-neighbor DNLS equation, in which the eigenvalues are
known to collide~\cite{Ka01,Ke09}.

\subsection{Two-site excitations}

Next we consider discrete solitons which are seeded, in the AL limit, by an
excitation localized on a pair of adjacent sites. According to the phase
difference between them, the emerging modes may be categorized as in-phase
(alias offsite-centered, alias symmetric) and out-of-phase (alias twisted
\cite{twisted}, i.e., antisymmetric) ones.

The results for the in-phase double-site mode\ produced at $\alpha =1.5$ are
represented in Fig.~\ref{Two}(a1) in the form of the $P(C)$ dependence.
Similarly to the single-site excitation, the curve cannot continue up to the
continuum limit, terminating at a critical value of the coupling constant $C$%
, which is the same as for the family of the single-site (onsite-centered)
modes. Profiles of the double-site discrete soliton are displayed, as a
function of $C$, in Fig.~\ref{Two}(a2), where the amplitude and width
increase with the growth of $C$. Like in the usual DNLS equation, the
transition which breaks the symmetry between the two sites leads to a
bifurcation of the translation eigenvalues from the origin to the real axis,
rendering the corresponding discrete solitons linearly unstable.\ In
contrast to the single-site excitation explored above, these eigenvalues
branch out from the origin, rather than from the edge of the continuous
spectrum, as can be explicitly seen in the AC\ limit. The variation of the
critical eigenvalues with respect to $C$ is presented in Figs.~\ref{Two}%
(a3,a4).

The in-phase two-site mode is made unstable by a real positive eigenvalue at
all values of $C\neq 0$. As $C$ increases further, a pair of eigenvalues
branches off from the edge of the continuous spectrum along the imaginary
axis to form a point spectrum. Simultaneously, the unstable eigenvalues are
decreasing in their magnitude. An example of an in-phase discrete soliton is
displayed in Fig.~\ref{Two}(b1) for $C=1$. Figure~\ref{Two}(b2) demonstrates
that that this discrete soliton is indeed linearly unstable, due to the
bifurcation from the origin, cf. Refs. \cite{Ka01,Pe05,Ke09}. Direct
simulations confirm the instability of this discrete soliton, see Figs.~\ref%
{Two}(b3,b4).

For the family of out-of-phase two-site (twisted) modes, the $P(C)$
dependence is presented in Fig.~\ref{Dipole}(a1), fixing $\alpha =1.5$. The
variation of critical eigenvalues is illustrated in Figs.~\ref{Dipole}%
(a2,a3). In this case, the translation eigenvalues bifurcate from the
origin, and first move along the imaginary axis. Then, at $C\approx 0.14$,
the collision with the edge of the continuous-spectrum band results in
transformation of pure imaginary stability eigenvalues into a complex
quartets. The latter effect is similar to the one known for the usual DNLS
equation. This observation is explained by the fact that these eigenvalues
have negative Krein signatures~\cite{Pe05,Ke09}, hence the collision gives
rise to a complex quartet, through the Hamilton-Hopf bifurcation~\cite%
{Va90,Ma91}. Thus, the family of the out-of-phase double-site discrete
solitons, unlike its in-phase counterpart, contains the stable segment, as
shown in Figs. \ref{Dipole}(a1,a2). A different phenomenon is that, as found
at $C\approx 0.6$, the eigenvalues bifurcating from the band edge of the
continuous spectrum touch the origin and thus become real unstable
eigenvalues. A stable discrete soliton is displayed in Fig.~\ref{Dipole}(b1)
for $C=0.1$, together with its spectrum of eigenvalues for small
perturbations, and stable propagation in the direct simulations, see Figs.~%
\ref{Dipole}(b2,b3). When the coupling constant increases to $C=0.2$, the
discrete soliton exhibited in Fig.~\ref{Dipole}(c1) is linearly unstable due
to complex eigenvalues, see Fig.~\ref{Dipole}(c2), as confirmed by the
direct simulations in Fig.~\ref{Dipole}(c3). A linearly unstable discrete
soliton is displayed in Fig.~\ref{Dipole}(d1) for $C=0.6$, with the
respective linear spectrum and unstable evolution exhibited in Figs.~\ref%
{Dipole}(d2) and (d3), respectively.

As said above, at $\alpha =2$ our DNLS equation with long-range
interactions, i.e., Eq.~(\ref{FNLSE-02v}), does not reduce to the usual
DNLS equation~(\ref{DNLS}) with the short-range interaction, making it necessary to
compare the two models. The results for different types of the discrete
solitons (onsite- and offsite-centered ones, as well as the twisted
solitons) are summarized in Fig.~\ref{Comp}, where one observes that both
models actually produce very close results.
\begin{figure*}[t]
\centering
\vspace{-0.1in} {\scalebox{0.68}[0.68]{\includegraphics{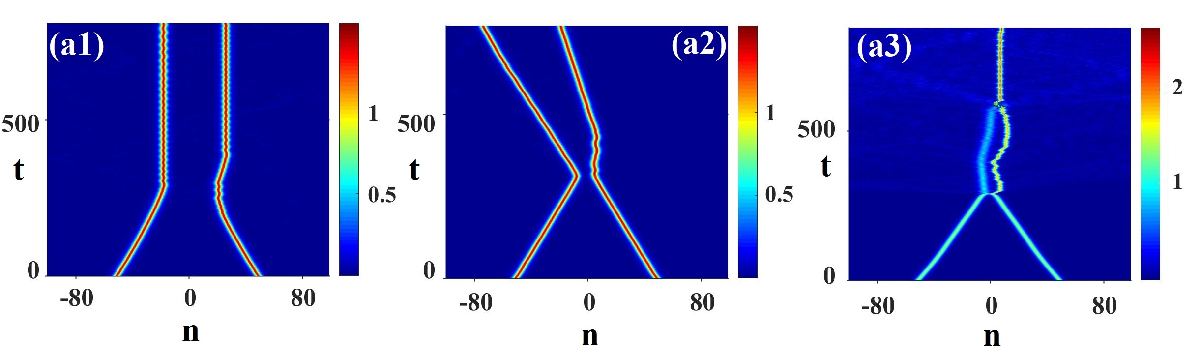}}}\hspace{-0.3in%
} \vspace{0.1in}
\caption{Collisions between one-site discrete solitons, initially located at
$n=\pm 50$, which are set in motion by kicks $k=\pm \protect\pi /64$ for $%
\protect\alpha =1.7$ (a1), $\protect\alpha =1.8$ (a2), and $\protect\alpha %
=2.0$ (a3). }
\label{Inter}
\end{figure*}
\section{Mobility and collisions of discrete solitons}

In addition to the breaking of the invariance with respect to spatial
displacements due to the discreteness, the Galilean invariance of the
present system is broken by the fractional diffraction~\cite{review,review2}%
, hence a nontrivial issue is mobility of initially kicked discrete
solitons. As is well known, discrete solitons may move across a lattice if
their kinetic energy exceeds the corresponding Peierls-Nabarro potential
barrier (PNPB), which is usually estimated as the free-energy~difference, $%
G=H-\omega P$, between the onsite- \ and offsite-centered modes
[corresponding to subscripts \textquotedblleft on" and \textquotedblleft
off" in Eq. (\ref{on-off})] for the same frequency $\omega $ or power $P$,
i.e.,
\begin{equation}
\Delta G=\Delta H-\omega \Delta P,  \label{FreeD}
\end{equation}%
where
\begin{equation}
\Delta H=H_{\text{off}}-H_{\text{on}},\quad \Delta P=P_{\text{off}}-P_{\text{%
on}}  \label{on-off}
\end{equation}%
with the subscripts denoting the two-site in-phase and single-site mode,
respectively~\cite{Kivshar,Cai,Morandotti,Me06,Ma08,Ke09}.
The so generated
estimate of the PNPB for the discrete soliton is presented in Fig.~\ref%
{Moving}(a) in the $\left( C,\alpha \right) $ plane. As expected, the free
energies of the onsite- and offsite-centered modes are close at larger $C$
(as one approaches the continuum limit), and, as LI $\alpha $ decreases, a
smaller difference of the free energies is observed for larger $C$. \ The
prediction produced by PNPB was verified by simulating Eq.~(\ref{FNLSE})
with the initial input in the form of the stationary discrete soliton $v_{n}$%
, to which the kick $k$ is applied:%
\begin{equation}
u_{n}(t=0)=v_{n}e^{ikn}.  \label{kick}
\end{equation}%
Several typical simulations are displayed in Figs.~\ref{Moving}(b1-b4) for $%
k=\pi /64$. In Fig.~\ref{Moving}(b1) one observes that the discrete soliton
cannot be set in motion by the kick at at $\alpha =1.5$. At $\alpha =1.7$
the application of the kick leaves the discrete soliton trapped after a
period of transient motion. Persistently moving discrete solitons are
observed in Figs.~\ref{Moving}(b3,b4) for $\alpha =1.9$ and $2.0$,
respectively, the motion being quicker in the latter case, which can be
explained by the PNPB estimate in Fig.~\ref{Moving}(a). Note that the moving
discrete solitons bounce back from the boundary of the simulation domain.

It is natural too to simulate collisions between mobile discrete solitons
moving in opposite directions, which are initialized by the input
\begin{equation}
u_{n}(0)=v_{n}^{+n_{0}}e^{ikn}+v_{n}^{-n_{0}}e^{-ikn},  \label{input}
\end{equation}%
where $v_{n}^{\pm n_{0}}$ are quiescent discrete solitons with centers
located at sites $n=\pm n_{0}$. Semi-elastic collisions are displayed in
Figs.~\ref{Inter}(a1,a2), for $\alpha =1.7$ and $1.8$, respectively. At $%
\alpha =2$, the collision is inelastic, as seen in Fig.~\ref{Inter}(a3).

Simulations clearly demonstrate that they lead to breaking of
the symmetry of the two-soliton set. As\ explained earlier in terms of a
continuum NLS equation with the non-fractional diffraction \cite{St23}, the
symmetry breaking in soliton-soliton collisions is not a spontaneous effect.
In fact, it is a predictable one, caused by a mismatch between the amplitude
and phase collision centers in the case of a soliton pair represented by a
complex wave field.

\section{Conclusions and discussions}

In the present work, we have proposed a novel form of the fractional
discrete nonlinear Schr\"{o}dinger equation, with L\'{e}vy index $\alpha \in
(0,2]$, based on the quasi-Riesz fractional derivative. In addition to the
fractional case, with $\alpha <2$, the case of $\alpha =2$ also provides a
new discrete nonlinear wave system with the long-range interaction decaying $\sim 1/l^{2}$
with the increase of distance $l$ between sites of the dynamical lattice. A
physical realization of the latter model is proposed, in terms of the array
of parallel chains of atoms or molecules carrying permanent magnetic or
electric dipole moments. The difference of the fractional
discrete NLS equation obtained in the limit of $\alpha =2$, with nonlocal
inter-site couplings, from the usual discrete equation with the
nearest-neighbor coupling is explained by the non-commutativity of the two
transitions: $\alpha \rightarrow 2$ and proceeding from the continuum setting to the
discrete lattice. The linear dispersion relation for linear waves in
this FDNLS system is obtained.

Then, starting from the anti-continuous limit, the
formation of discrete single- and double-site discrete solitons is
investigated. The stability of the discrete solitons is examined by the
calculation of eigenvalues for small perturbations, and is verified by means
of direct simulations. Moreover, based on the free-energy estimates of the PNPB
(Peierls-Nabarro barrier), mobility of the discrete solitons is predicted
and tested in direct simulations. The simulations are also used to study
collisions between the moving discrete solitons.

The present study suggests directions for further work. One aspect is that
discrete solitons do not persist in the continuum limit due to long-range
interaction. A detailed analysis of how the existence of the discrete
solitons terminates as the continuum limit is approached would be relevant
to perform. As briefly mentioned above, the present discrete system with the
long-range intersite interactions may be naturally generalized by
considering spatially periodic modulation of the interactions.

Another natural option is to consider nonlinear modes of other types, such
as dark solitons~\cite{Ki94,Fi07}, surface waves~\cite{Ma05}, etc. A
challenging extension is to introduce a 2D version of the system and address
modes such as 2D fundamental and vortex solitons. Further, as concerns
physically relevant 2D settings, an intriguing possibility is to introduce a
semi-discrete system (cf. Ref.~\cite{Zh19}) described by the fractional NLS
equation
\begin{equation}
\begin{array}{rl}
i\partial _{z}u_{n}= & \displaystyle\frac{1}{2}\left[ \left( -\frac{\partial
^{2}}{\partial x^{2}}\right) ^{\alpha /2}+\left( -\frac{\widehat{\partial }^{2}}{\widehat{%
\partial}n^{2}}\right) ^{\beta /2}\right] u_{n}\vspace{0.1in}
\\
& \displaystyle+F(x,n,\left\vert u_{n}\right\vert ^{2})u_{n},%
\end{array}%
\end{equation}%
where $x$ and $n$ are continuous and discrete coordinates, respectively,
with the corresponding L\'{e}vy indices $\alpha $ and $\beta $, the
continuous and discrete fractional derivatives are defined, severally, as
per Eqs. (\ref{Riesz derivative}) and (\ref{Riesz}), and $F$ represents a
general nonlinear term. Such studies are currently in progress and will be
reported elsewhere.

\section*{Acknowledgments}


The work of Z.Y. was supported by the National Natural Science Foundation of
China (No. 11925108). The work of B.A.M. is supported, in part, by grant No.
1695/22 from the Israel Science Foundation.


\begin{thebibliography}{99}
\bibitem{Uchaikin} V. V. Uchaikin, \emph{Fractional Derivatives for
Physicists and Engineers} (Springer, New York, 2013).

\bibitem{Caputo} M. Caputo, Linear model of dissipation whose Q is almost
frequency independent. II, Geophys. J. Inter. \textbf{13}, 529--539 (1967).

\bibitem{fd1} M. F. Shlesinger, G. M. Zaslavsky, and J. Klafter, Strange
kinetics, Nature (London) \textbf{363}, 31 (1993).

\bibitem{fd2} N. C. Petroni and M. Pusterla, L\'evy processes and
Schr\"odinger equation, Physica A \textbf{388}, 824 (2009).

\bibitem{Lask1} N. Laskin, Fractional quantum mechanics, Phys. Rev. E
\textbf{62}, 3135 (2000).

\bibitem{Lask2} N. Laskin, Fractional quantum mechanics and L\'{e}vy path
integrals, Phys. Lett. A \textbf{268}, 298-305 (2000).

\bibitem{Lask3} N. Laskin, Fractional Schr\"odinger equation, Phys. Rev. E
\textbf{66}, 056108 (2002).

\bibitem{fd-na} P. Barthelemy, J. Bertolotti, and D. S. Wiersma, A L\'evy
flight for light, Nature \textbf{453}, 495-8 (2008).

\bibitem{fd-np} N. Mercadier, W. Guerin, M. Chevrollier, and R. Kaiser,
L\'evy flights of photons in hot atomic vapours, Nat. Phys. \textbf{5},
602-605 (2009).

\bibitem{St13} B.A. Stickler, Potential condensed-matter realization of
space-fractional quantum mechanics: the one-dimensional L\'evy crystal,
Phys. Rev. E \textbf{88}, 012120 (2013).

\bibitem{fd3} M. I. Molina, Fractionality and PT symmetry in a square
lattice, Phys. Rev. A \textbf{106}, L040202 (2022).

\bibitem{Ku22} S. Kumar, P. Li, and B. A. Malomed, Domain walls in
fractional media, Phys. Rev. E \textbf{106}, 054207 (2022).

\bibitem{nG} G. Samorodnitsky and M. S. Taqqu, \textit{Stable Non-Gaussian
Random Processes} (Chapman and Hall, New York, 1994).

\bibitem{Lask4} N. Laskin, \emph{Fractional Quantum Mechanics} (World
Scientific: Singapore, 2018).

\bibitem{b1} S. G. Samko, A. A. Kilbas, and I. O. Marichev, \textit{%
Fractional Integrals and Derivatives: Theory and Applications} (Gordon and
Breach, 1993).

\bibitem{b3} R. Hilfer, \textit{Applications of fractional calculus in
physics} (World Scientific, Singapore, 2000).

\bibitem{b2} C. A. Monje, Y. Q. Chen, B. M. Vinagre, D. Xue, and V. Feliu,
\textit{Fractional-order systems and controls} (Springer, Berlin, 2010).

\bibitem{b4} V. E. Tarasov, \textit{Fractional dynamics: applications of
fractional calculus to dynamics of particles, fields and media} (Springer,
Berlin 2011).

\bibitem{b5} V. Uchaikin and R. Sibatov, \textit{Fractional kinetics in
solids: Anomalous charge transport in semiconductors, dielectrics and
nanosystems} ( World Scientific Publishing, Singapore, 2013).

\bibitem{PR1} R. Metzler and J. Klafter, The random walk's guide to
anomalous diffusion: a fractional dynamics approach, Phys. Rep. \textbf{339}%
, 1-77 (2000).

\bibitem{PR2} G. M. Zaslavsky, Chaos, fractional kinetics, and anomalous
transport, Phys. Rep. \textbf{371}, 461 (2002).


\bibitem{fd-rev18} H. Sun, Y. Zhang, D. Baleanu, W. Chen, and Y. Chen, A new
collection of real world applications of fractional calculus in science and
engineering, Commun. Nonlinear Sci. Numer. Simulat. \textbf{64}, 213-231
(2018).

\bibitem{GuoXu} X. Guo and M. Xu, Some physical applications of fractional
Schr\"{o}dinger equation, J. Math. Phys. \textbf{47}, 082104 (2006).

\bibitem{Mandelbrot} B. B. Mandelbrot, \emph{The Fractal Geometry of Nature}
(W. H. Freeman, New York, 1982).

\bibitem{Feynman} S. Albeverio, R. Hoegh-Krohn, and S. Mazzucchi, \emph{%
Mathematical Theory of Feynman Path Integrals : An Introduction} (Springer,
Berlin and Heidelberg, 2008).

\bibitem{Longhi} S. Longhi, Fractional Schr\"{o}dinger equation in optics,
Opt. Lett. \textbf{40}, 1117-1120 (2015).

\bibitem{Zhang15} Y. Zhang, X. Liu, M. R. Beli\'c, W. Zhong, Y. Zhang, and
M. Xiao, Propagation dynamics of a light beam in a fractional Schr\"odinger
equation, Phys. Rev. A \textbf{115}, 180403 (2015).

\bibitem{Zhong16} W. P. Zhong, M. R. Beli\'c, B. A Malomed, Y. Zhang, and T.
Huang, Spatiotemporal accessible solitons in fractional dimensions, Phys.
Rev. E \textbf{94}, 012216 (2016).

\bibitem{ZZ16} Y. Zhang, H. Zhong, M. R. Beli\'c, Y. Zhu, W. Zhong, Y.
Zhang, D. N. Christodoulides, and M. Xiao, PT-symmetry in a fractional
Schr\"odinger equation, Laser Photonics Rev. \textbf{10}, 526-531 (2016).

\bibitem{Riesz} M. Cai and C. P. Li, On Riesz derivative, Fractional Cal.
Appl. Anal. \textbf{22}, 287-301 (2019).

\bibitem{review} B. A. Malomed, Optical solitons and vortices in fractional
media: A mini-review of recent results, Photonics \textbf{8}, 353 (2021).

\bibitem{review2} B. A. Malomed, Basic fractional nonlinear-wave models and
solitons, Chaos, in press.

\bibitem{bec-fd} V. A. Stephanovich , E. V. Kirichenko , G. Engel , and A.
Sinner, Spin-orbit-coupled fractional oscillators and trapped Bose-Einstein
condensates, Phys. Rev. E \textbf{109}, 014222 (2024).

\bibitem{Pit-Str} L. P. Pitaevskii and S. Stringari, \emph{Bose-Einstein
Condensation} (Oxford University Press, Oxford, 2003).

\bibitem{Huang08} C. Huang, H. Deng, W. Zhang, F. Ye, and L. Dong,
Fundamental solitons in the nonlinear fractional Schr\"odinger equation with
a PT-symmetric potential, EPL \textbf{122}, 24002 (2008).


\bibitem{Huang16} C. Huang and L. Dong, Gap solitons in the nonlinear
fractional Schr\"odinger equation with an optical lattice, Opt. Lett.
\textbf{41}, 5636-5639 (2016).

\bibitem{Zhang16} L. Zhang, C. Li, H. Zhong, C. Xu, D. Lei, Y. Li, and D.
Fan, Propagation dynamics of super-Gaussian beams in fractional
Schr\"odinger equation: from linear to nonlinear regimes, Opt. Exp. \textbf{%
24}, 14406-14418 (2016).

\bibitem{Yao18} X. Yao and X. Liu, Off-site and on-site vortex solitons in
space-fractional photonic lattices, Opt. Lett. \textbf{43}, 5749-5752 (2018).

\bibitem{Guo18} M. Chen, S. Zeng, D. Lu, W. Hu, and Q. Guo, Optical
solitons, self-focusing, and wave collapse in a space-fractional
Schr\"odinger equation with a Kerr-type nonlinearity, Phys. Rev. E \textbf{98%
}, 022211 (2018).

\bibitem{Zeng20} L. Zeng and J. Zeng, Preventing critical collapse of
higher-order solitons by tailoring unconventional optical diffraction and
nonlinearities, Commun. Phys. \textbf{3}, 26 (2020).

\bibitem{Xie19} J. Xie, X. Zhu, and Y. He, Vector solitons in nonlinear
fractional Schr\"{o}dinger equations with parity-time-symmetric optical
lattices, Nonlinear Dyn. \textbf{97}, 1287 (2019).

\bibitem{Li20} P. Li, B. A. Malomed, and D. Mihalache, Symmetry-breaking
bifurcations and ghost states in the fractional nonlinear Schr\"odinger
equation with a PT-symmetric potential, Opt. Lett. \textbf{46}, 3267 (2021).

\bibitem{Zhong23} M. Zhong, L. Wang, P. Li, and Z. Yan, Spontaneous symmetry
breaking and ghost states supported by the fractional PT-symmetric saturable
nonlinear Schr\"odinger equation, Chaos \textbf{33}, 013106 (2023).

\bibitem{Zhong24} M. Zhong, Y. Chen, Z. Yan, and B. A. Malomed, Suppression
of soliton collapses, modulational instability and rogue-wave excitation in
two-L\'evy-index fractional Kerr media, Proc. R. Soc. A \textbf{480},
20230765 (2024).

\bibitem{proceedings} C. Klein, C. Sparber, and P. Markowich, Numerical
study of fractional nonlinear Schr\"{o}dinger equations, Proc. R. Soc. A
\textbf{470}, 20140364 (2014).

\bibitem{KA} Y. S. Kivshar and G. P. Agrawal, \emph{Optical Solitons: From
Fibers to Photonic Crystals} (Academic Press, San Diego, 2003).

\bibitem{Shilong} S. Liu, Y. Zhang, B. A. Malomed, and E. Karimi,
Experimental realisations of the fractional Schr\"{o}dinger equation in the
temporal domain, Nature Comm. \textbf{14}, 222 (2023).

\bibitem{PRE95} Y. B. Gaididei, S. F. Mingaleev,  P. L. Christiansen,  and K. $\emptyset$. Rasmussen,
 Effects of nonlocal dispersive interactions on self-trapping excitations,  Phys. Rev. E \textbf{55}, 6141 (1997).




\bibitem{CMP17} M. Jenkinson, and M. I. Weinstein,   Discrete solitary waves in systems with nonlocal interactions and the Peierls-Nabarro barrier,
	Commun. Math. Phys.  \textbf{351}, 45-94 (2017).

\bibitem{mathematical} O. Ciaurri, L. Roncal, P. R. Stinga, J. L. Torrea,
and J. L. Varona, Nonlocal discrete diffusion equations and the fractional
discrete Laplacian, regularity and applications, \textnormal{Adv. Math.}
\textbf{330}, 688-738 (2018).

\bibitem{Molina} M. I. Molina, The fractional discrete nonlinear Schr\"{o}%
dinger equation, Phys. Lett. A \textbf{384}, 126180 (2020).

\bibitem{Molina2} M. I. Molina, The two-dimensional fractional discrete
nonlinear Schr\"{o}dinger equation, Phys. Lett. A \textbf{384}, 126835
(2020).

\bibitem{Molina-electro} M. I. Molina, Fractional nonlinear electrical
lattice, Phys. Rev. E \textbf{104}, 024219 (2021).

 \bibitem{Ta06} V. E. Tarasov,  Continuous limit of discrete systems with long-range interaction,
 J. Phys. A: Math. Gen.  \textbf{39}, 14895 (2006).

 \bibitem{Ta062} V. E. Tarasov,  Map of discrete system into continuous,
  J. Math. Phys. \textbf{47} 092901 (2006).

 \bibitem{Ta16} V. E. Tarasov, Exact discretization by Fourier transforms,
 Commun. Nonlinear Sci. Numer. Simul. \textbf{37}, 31-61 (2016).


\bibitem{Ke09} P. G. Kevrekidis, \textit{The discrete nonlinear
Schr\"odinger equation: mathematical analysis, numerical computations and
physical perspectives} (Springer, Berlin, 2009).

\bibitem{dnls} J. C. Eilbeck, P. S. Lomdahl, and A. C. Scott, The discrete
selftrapping equation, Physica D \textbf{16}, 318-338 (1985).



\bibitem{Le08} F. Lederer, G. I. Stegeman, D. N. Christodoulides, G.
Assanto, M. Segev, and Y. Silberberg Discrete solitons in optics, Phys. Rep.
\textbf{463}, 1-126 (2008).

\bibitem{TLN} L. N. Trefethen, {\it Spectral methods in MATLAB} (SIAM, 2000).

\bibitem{At91} K. Atkinson, \textit{An Introduction to Numerical Analysis}
(John Wiley \& Sons, New York, 1991).

\bibitem{dipolar-BEC} T. Lahaye, C. Menotti, L. Santos, M. Lewenstein, and
T. Pfau, The physics of dipolar bosonic quantum gases, Rep. Prog. Phys.
\textbf{72}, 126401 (2009).

\bibitem{HS} H. Sakaguchi and B. A. Malomed, Suppression of the
quantum-mechanical collapse by repulsive interactions in a quantum gas,
Phys. Rev. A \textbf{83}, 013607 (2011).

\bibitem{Ma94} R. S. MacKay, and S. Aubry, Proof of existence of breathers
for time-reversible or Hamiltonian networks of weakly coupled oscillators,
Nonlinearity \textbf{7}, 1623 (1994).

\bibitem{Au97} S. Aubry, Breathers in nonlinear lattices: Existence, linear
stability and quantization, Physica D \textbf{103}, 201-250 (1997).

\bibitem{He99} D. Hennig and G. P. Tsironis, Wave transmission in nonlinear
lattices, \textnormal{Phys. Rep.} \textbf{307}, 333-432 (1999).

\bibitem{Du00} H. R. Dullin and J. D. Meiss, Generalized H\'enon maps: the
cubic diffeomorphisms of the plane, \textnormal{Physica D} \textbf{143},
262-289 (2000).

\bibitem{Ka01} T. Kapitula and P. Kevrekidis, Stability of waves in discrete
systems, \textnormal{Nonlinearity} \textbf{14}, 533 (2001).

\bibitem{Al04} G. L. Alfimov, V. A. Brazhnyi, and V. V. Konotop, On
classification of intrinsic localized modes for the discrete nonlinear
Schr\"odinger equation, Physica D \textbf{194}, 127-150 (2004).

\bibitem{Pe05} D. E. Pelinovsky, P. G. Kevrekidis, and D. J. Frantzeskakis,
Stability of discrete solitons in nonlinear Schr\"odinger lattices,
\textnormal{Physica D} \textbf{212}, 1-19 (2005).

\bibitem{Le92} H. Levy and F. Lessman, \textit{Finite Difference Equations}
(Courier Corporation, 1992).

\bibitem{Ho12} R. A. Horn and C. R. Johnson, \textit{Matrix Analysis}
(Cambridge University Press, Cambridge, 2012).

\bibitem{Ca23} J. Catarecha, J. Cuevas-Maraver, and P. G. Kevrekidis,
\textnormal{Breathers in the fractional Frenkel-Kontorova model},
arXiv:2311.01809 (2023).

\bibitem{Fl98} S. Flach, Breathers on Lattices with Long Range Interaction,
\textnormal{\ Phys. Rev. E} \textbf{58}, R4116 (1998).


\bibitem{St04} M. Stepi\'c, D. Kip, L. Hadzievski, and A. Maluckov,
One-dimensional bright discrete solitons in media with saturable
nonlinearity, Phys. Rev. E \textbf{69}, 066618 (2004).

\bibitem{Ha04} L. Hadzievski, A. Maluckov, M. Stepi\'{c}, and D. Kip, Power
controlled soliton stability and steering in lattices with saturable
nonlinearity, Phys. Rev. Lett. \textbf{93}, 033901 (2004).

\bibitem{Me06} T. R. Melvin, A. R. Champneys, P. G. Kevrekidis, and J.
Cuevas, Radiationless traveling waves in saturable nonlinear Schr\"odinger
lattices, Phys. Rev. Lett. \textbf{97}, 124101 (2006).

\bibitem{Jo00} M. Johansson, and S. Aubry, Growth and decay of discrete
nonlinear Schr\"odinger breathers interacting with internal modes or
standing-wave phonons, Phys. Rev. E \textbf{61}, 5864 (2000).

\bibitem{Va90} J. C. van der Meer, Hamiltonian Hopf bifurcation with
symmetry, Nonlinearity \textbf{3}, 1041 (1990).

\bibitem{Ma91} R. S. MacKay, Movement of eigenvalues of Hamiltonian
equilibria under non-Hamiltonian perturbation, Phys. Lett. A \textbf{155},
266-268 (1991).

\bibitem{twisted} S. Darmanyan, A. Kobyakov, and F. Lederer, Stability of
strongly localized excitations in discrete media with cubic nonlinearity,
Zh. Eksp. Teor. Fiz. 113, 1253-1260 (1998) [J. Exp. Theor. Phys. \textbf{86}%
, 682-686 (1998)].



\bibitem{Kivshar} Y. S. Kivshar and D. K. Campbell, Peierls-Nabarro
potential barrier for highly localized nonlinear modes, Phys. Rev. E \textbf{%
48}, 3077-3081 (1993).

\bibitem{Cai} D. Cai, A. R. Bishop, and N. Gr\o nbech-Jensen, Localized
states in discrete nonlinear Schr\"{o}dinger equation, Phys. Rev. Lett.
\textbf{72}, 591-595 (1994).

\bibitem{Morandotti} R. Morandotti,, U. Peschel, J. S. Aitchison, H. S.
Eisenberg, and Y. Silberberg, Dynamics of discrete solitons in optical
waveguide arrays, Phys. Rev. Lett. \textbf{83}, 2726-2729 (1999).

\bibitem{Ma08} A. Maluckov, L. Hadzievski, and B. A. Malomed, Staggered and
moving localized modes in dynamical lattices with the cubic-quintic
nonlinearity, Phys. Rev. E \textbf{77}, 036604 (2008).

\bibitem{St23} L. Khaykovich and B. A. Malomed, Deviation from one
dimensionality in stationary properties and collisional dynamics of
matter-wave solitons, Phys. Rev. A \textbf{74}, 023607 (2006).

\bibitem{Ki94} Y. S. Kivshar, W. Kr\'{o}likowski, and O. A. Chubykalo, Dark
solitons in discrete lattices, Phys. Rev. E \textbf{50}, 5020 (1994).

\bibitem{Fi07} E. P. Fitrakis, P. G. Kevrekidis, H. Susanto, and D. J.
Frantzeskakis, Dark solitons in discrete lattices: Saturable versus cubic
nonlinearities, Phys. Rev. E \textbf{75}, 066608 (2007).

\bibitem{Ma05} K. G. Makris, S. Suntsov, D. N. Christodoulides, G. I.
Stegeman, and A. Hache, Discrete surface solitons, Opt. Lett. \textbf{30},
2466-2468 (2005).

\bibitem{Zh19} X. Zhang, X. Xu, Y. Zheng, Z. Chen, B. Liu, C. Huang, B. A.
Malomed, and Y. Li, Semidiscrete quantum droplets and vortices, Phys. Rev.
Lett. \textbf{123}, 133901 (2019).
\end{thebibliography}
\end{document}